\documentclass[sort&compress,dvipdfmx]{elsarticle} % dvipdfmx to avoid the overlap of the figures and caption

\usepackage{lineno}
\usepackage{multirow}

\usepackage{amsfonts}
\usepackage{graphicx}
\usepackage{latexsym}
\usepackage{amsmath,amssymb}
\usepackage{mathrsfs}
\usepackage{hhline}
\usepackage{dcolumn}
\usepackage{bm}
\usepackage[usenames,dvipsnames]{xcolor}
\usepackage{supertabular}
\usepackage{multirow}

\usepackage{bigstrut}

\usepackage{esvect} % \vv

\usepackage[normalem]{ulem} % strike out

\newcolumntype{Y}{>{\raggedleft\arraybackslash}p{15mm}} 

\usepackage{url}

\begin{document}

\title{Clustering coefficients for correlation networks}

\author{Naoki Masuda}
\ead{naoki.masuda@bristol.ac.uk}
\address{Department of Engineering Mathematics, University of Bristol, Bristol, United Kingdom}

\author{Michiko Sakaki}
\address{School of Psychology and Clinical Language Sciences, University of Reading, Earley Gate, Whiteknights Road, Reading, United Kingdom}
\address{Research Institute, Kochi University of Technology, Kami, Kochi, Japan}

\author{Takahiro Ezaki}
\address{PRESTO, JST, 4-1-8 Honcho, Kawaguchi, Saitama, Japan}

\author{Takamitsu Watanabe}
\address{Institute of Cognitive Neuroscience, University College London, 17 Queen Square, London, WC1N 3AZ, United Kingdom}

\begin{abstract}
Graph theory is a useful tool for deciphering structural and functional networks of the brain on various spatial and temporal scales. The clustering coefficient quantifies the abundance of connected triangles in a network and is a major descriptive statistics of networks. For example, it finds an application in the assessment of small-worldness of brain networks, which is affected by attentional and cognitive conditions, age, psychiatric disorders and so forth. However, it remains unclear how the clustering coefficient should be measured in a correlation-based network, which is among major representations of brain networks. In the present article, we propose clustering coefficients tailored to correlation matrices. The key idea is to use three-way partial correlation or partial mutual information to measure the strength of the association between the two neighbouring nodes of a focal node relative to the amount of pseudo-correlation expected from indirect paths between the nodes. Our method avoids the difficulties of previous applications of clustering coefficient (and other) measures in defining correlational networks, i.e., thresholding on the correlation value, discarding of negative correlation values, the pseudo-correlation problem and full partial correlation matrices whose estimation is computationally difficult. For proof of concept, we apply the proposed clustering coefficient measures to functional magnetic resonance imaging data obtained from healthy participants of various ages and compare them with conventional clustering coefficients. We show that the clustering coefficients decline with the age. The proposed clustering coefficients are more strongly correlated with age than the conventional ones are. We also show that the local variants of the proposed clustering coefficients (i.e., abundance of triangles around a focal node) are useful in characterising individual nodes. In contrast, the conventional local clustering coefficients were strongly correlated with and therefore may be confounded by the node's connectivity. The proposed methods are expected to help us to understand clustering and lack thereof in correlational brain networks, such as those derived from functional time series and across-participant correlation in neuroanatomical properties.
\end{abstract}

\maketitle

% \linenumbers

\section{Introduction\label{sec:introduction}}

Networks have been proven to be a useful language to understand structural and functional properties of the brain. The research field is collectively called network neuroscience \cite{Bassett2017NatNeurosci}. Initial studies in network neuroscience revealed that brain networks on various spatial scales have properties common to other biological and non-biological networks, such as the small-world property and community structure. More recent studies tend to depend on the availability of new tools to record data with, look at other properties of brain networks such as network hubs, rich clubs and economic efficiency, and  endeavour into the analysis of impaired brains \cite{Bullmore2009NatRevNeurosci,Sporns2011book,Fornito2013Neuroimage,Stam2014NatRevNeurosci,Bassett2017NatNeurosci}.

In this article, we focus on a measure which has often been applied to brain (and other) networks: clustering coefficient \cite{Watts1998Nature}. The clustering coefficient quantifies the abundance of connected triangles in a network. In network neuroscience, the clustering coefficient has been shown to be a useful quantity for understanding function-structure associations in the brain for at least the following two reasons. First, it is one of the two building blocks with which to measure the small-worldness of a network; small-world networks are those having a large clustering coefficient and a small shortest path length between two nodes (such as regions of interest; ROIs) on average \cite{Watts1998Nature,Bullmore2009NatRevNeurosci}. Brain networks are usually small-world networks in  this sense \cite{Achard2006JNeurosci,Bassett2006Neuroscientist}. Loss of small-worldness is a signature of, for example, Alzheimer disease \cite{Supekar2008PlosComputBiol,Brier2014NeurobiolAging} and schizophrenia \cite{LiuLiang2008Brain}. Second, the abundance of connected triangles around a given node, which is measured by local variants of the clustering coefficient, informs us of other structure and functions of networks, namely, community structure \cite{Radicchi2004PNAS,Palla2005Nature} and local efficiency \cite{Latora2001PhysRevLett}. Both community structure and local efficiency are often measured for brain networks \cite{Achard2007PlosComputBiol,Bullmore2009NatRevNeurosci,Rubinov2010Neuroimage,Rubinov2011Neuroimage};
for example, community structure of functional brain networks is less pronounced in childhood-onset schizophrenia than controls \cite{Alexanderbloch2010FrontSystNeurosci}.

However, the current measurement of the clustering coefficient can be easily fooled when it is applied to correlational brain/neuronal networks, where the connectivity between two nodes is defined by Pearson correlation and potentially some other correlation indices. Such correlational brain networks are often built on the basis of a correlation measure between two ROIs such as the pairwise correlation between time-dependent blood oxygen level-dependent (BOLD) signals obtained from functional magnetic resonance imaging (fMRI) or neural signals obtained from electroencephalogram (EEG) and magnetoencephalogram (MEG) \cite{Bullmore2009NatRevNeurosci,Bassett2017NatNeurosci}. Correlational networks are also employed to construct structural networks of the brain, where an edge between two ROIs is calculated as the across-participant correlation in the cortical thickness \cite{Alexanderbloch2013NatRevNeurosci,Evans2013Neuroimage}.
A naive application of network analysis tools, including the clustering coefficient, to such correlation networks can go awry due to the following reasons.

First, a network derived from a correlation matrix tends to have many triangles owing to the so-called indirect paths, i.e., a correlation between nodes $i$ and $j$ and one between $i$ and $\ell$ result in a correlation between $j$ and $\ell$ even when there is no direct relationship between $j$ and $\ell$ \cite{Adachi2012CerebCortex,Zalesky2012Neuroimage}. This mathematical property raises the clustering coefficient values. The same pseudo-correlation effect also automatically produces 
an inflated correlation between the connectivity of node $i$ and the local clustering coefficient (i.e., which refers to the abundance of triangles around a particular node $i$ and has been used for characterising individual ROIs
\cite{SpornsZwi2004Neuroinfo,Achard2006JNeurosci,He2007CerebCortex,Alexanderbloch2010FrontSystNeurosci,Lynall2010JNeurosci,Power2010Neuron,Vandenheuvel2010JNeurosci,Vandenheuvel2011JNeurosci,Wee2011Neuroimage,Fornito2012Neuroimage,Tijms2013NeurobiolAging,Salallonch2014NeurobiolAging}), as we will show (section~\ref{sub:local C}). One remedy is to use appropriate null models \cite{Zalesky2012Neuroimage}, which respect the natural constraints imposed on correlation matrices including a large clustering coefficient value even in the case of networks generated at random. Nevertheless, this solution does not address the issue of the threshold value, which we will discuss below.
The partial correlation matrix is a method of choice for removing pseudo-correlation between ROIs that is present in networks based on the Pearson correlation matrix. However, estimation of the partial correlation matrix is difficult, particularly when the number of image volumes is relatively small as compared to the number of ROIs, which is typical of fMRI experiments \cite{Schafer2005StatApplGenetMolBiol,Ryali2012Neuroimage,Brier2015Neuroimage}.

Second, to create a network, we conventionally threshold on the correlation value to dichotomise the presence or absence of an edge between each pair of ROIs. However, the choice of the threshold is arbitrary \cite{Rubinov2010Neuroimage,Rubinov2011Neuroimage,Devikofallani2014PhilTransRSocB,Garrison2015Neuroimage} and results of graph-theoretical analyses often depend on the choice of the threshold \cite{Zalesky2012Neuroimage,Garrison2015Neuroimage,Jalili2016SciRep}.
Specifically, clustering coefficient values considerably depend on the threshold value \cite{Zalesky2012Neuroimage,Garrison2015Neuroimage}.
One can avoid thresholding by using weighted networks, i.e., networks with weighted edges \cite{Rubinov2010Neuroimage,Rubinov2011Neuroimage}. There are several definitions of clustering coefficient for weighted networks \cite{Barrat2004PNAS,Onnela2005PhysRevE,ZhangHorvath2005StatApplGenetMolBiol,Saramaki2007PhysRevE,Rubinov2010Neuroimage,Rubinov2011Neuroimage,Costantini2014PlosOne,WangGhumare2017NeuralComput}. However, it is unclear how the weighted network approach should deal with negatively weighted edges; most network analysis tools including the clustering coefficient assume non-negative edges \cite{Newman2010book}. An interesting possibility is to separately analyse networks composed of positive edges and those composed of negative edges, and then to combine the measurements obtained from the two types of networks \cite{Rubinov2011Neuroimage}. However, there seems to be no consensus regarding the treatment of negatively signed edges \cite{Sporns2016AnnuRevPsychol}.

In the present study, we develop two clustering coefficients tailored to correlation matrices. The first type of clustering coefficient is based on three-way partial correlation coefficient. The second type is based on partial mutual information. Partial mutual information is a nonlinear correlation measure, which is defined as the conventional mutual information between two random variables but conditioned on other variables \cite{Frenzel2007PhysRevLett}. These clustering coefficients are expected to overcome some of the aforementioned difficulties. First, they discount the effect of indirect paths to quantify association between two neighbours of a node given the activity of the focal node. In this manner, we avoid both the problem of pseudo-correlation in ordinary correlation matrices and computational issues in the calculation of partial correlation matrices. Second, as in the case of the clustering coefficients for weighted networks, our clustering coefficients do not use thresholding on the correlation value. Third, we measure how far the realised pairwise correlation value is  (no matter positive or negative) from the correlation anticipated by the presence of indirect paths. Although this treatment does not solve the problem of negative edges, we intend to use the information contained in negative as well as positive edges in this manner. For a proof of concept, we apply the proposed clustering coefficient indices to fMRI data obtained from healthy subjects with a wide range of age. We show that the clustering coefficients are negatively correlated with the age. This observation is in general less pronounced with the conventional clustering coefficient measures, although decline in the clustering coefficient with ageing should not be regarded as a ground truth in light of the literature \cite{WangLi2010Neuroimage,Matthaus2012BrainConn,ZhuWen2012NeurobiolAging,Brier2014NeurobiolAging,LiuKe2014PlosOne,Salallonch2014NeurobiolAging,Knyazev2015NeurobiolAging,Grady2016NeurobiolAging}. We also show that the local clustering coefficients at specific ROIs provide information orthogonal to the mere strength of connectivity and that their association with the participant's age is independent of brain systems.

\section{Methods}

\subsection{Functional connectivity}
 
We used $N_{\rm ROI}=30$ regions of interest (ROIs) whose coordinates were determined in a
previous study \cite{Fair2009PlosComputBiol}. Note that we excluded the four cerebellar ROIs out of the 34 ROIs. The system of the 30 ROIs contained the default mode network (DMN; 12 ROIs), cingulo-opercular network (CON; 7 ROIs) and fronto-parietal network (FPN; 11 ROIs).

Denote by $\rho(i,j)$ the Pearson correlation coefficient between the BOLD signals at two ROIs $i$ and $j$ ($1\le i, j\le N_{\rm ROI}$). We primarily use $\rho(i,j)$ as a measure of functional connectivity between ROIs. However, we will discount the effect of indirect paths, which is present when the edge between ROIs $i$ and $j$ is solely determined by $\rho(i,j)$, by defining new clustering coefficients (section~\ref{sub:cor A and M}).

For comparison purposes, we will also examine conventional clustering coefficients for networks (sections~\ref{sub:unweighted C} and \ref{sub:weighted C}), which are applied to the Pearson correlation matrix and the partial correlation matrix. The partial correlation matrix, which we use as a benchmark, is 
an alternative measure of functional connectivity \cite{Salvador2005CerebCortex,Marrelec2006Neuroimage}, and its $(i, j)$ element is estimated by $\overline{\rho}^{\rm partial}(i, j) =-\text{cov}^{-1}(i,j)/\sqrt{\text{cov}^{-1}(i,i)\text{cov}^{-1}(j,j)}$, where $\text{cov}$ denotes the covariance matrix \cite{Whittaker1990book}. 
It should be noted that $\rho(i,j)=\rho(j,i)$ and $\overline{\rho}^{\rm partial}(i,j) = \overline{\rho}^{\rm partial}(j,i)$. We interchangeably use node and ROI in the following.

\subsection{Average functional connectivity}

We used the following two indices of average functional connectivity: the pairwise Pearson correlation coefficient averaged over all pairs of ROIs, denoted by $s$, and the same average but only over the ROI pairs having the non-negative $\rho(i, j)$ values, denoted by $s^+$. The introduction of $s^+$ is motivated by the observation that the interpretation of negative correlation coefficients remains difficult \cite{Fox2009JNeurophysiol,Murphy2009Neuroimage,Rubinov2011Neuroimage,Fornito2013Neuroimage}.

\subsection{Clustering coefficients for unweighted networks\label{sub:unweighted C}}

In this section and the next, we explain the previously proposed clustering coefficients for unweighted and weighted networks based on the Pearson correlation coefficient, $\rho(i,j)$. Those based on the partial correlation coefficient, $\overline{\rho}^{\rm partial}(i,j)$, are analogously calculated.

To construct an unweighted functional network, we lay an edge between nodes $i$ and $j$ ($1\le i \neq j\le N$) if and only if $\rho(i,j)\ge \theta$, where $\theta$ is a pre-determined threshold. The generated network is undirected. We denote the adjacency matrix of the network by $A=(a_{ij})$, where $1\le i, j\le N_{\rm ROI}$. In other words, $a_{ij}=1$ if $(i, j)$ is an edge and $a_{ij}=0$ otherwise. The clustering coefficient represents the abundance of connected triangles in a network \cite{Watts1998Nature}.
The local clustering coefficient of node $i$ is defined by
\begin{align}
C_i^{\rm unw} =& \frac{(\text{Number of connected triangles including node $i$})}
{k_i (k_i-1)/2}\notag\\
=& \frac{\sum_{\substack{1\le j < \ell \le N_{\rm ROI}\\ j, \ell\neq i}} a_{ij}a_{i\ell}a_{j\ell}}
{k_i (k_i-1)/2},
\label{eq:C_i}
\end{align}
where $k_i = \sum_{j=1}^{N_{\rm ROI}} a_{ij} = \sum_{j=1}^{N_{\rm ROI}} a_{ji}$ is the degree of node $i$, i.e., the number of edges to which node $i$ is adjacent. The denominator on the right-hand side of Eq.~\eqref{eq:C_i} represents the largest possible number of triangles to which node $i$ belongs. Note that $0\le C_i^{\rm unw}\le 1$ ($1\le i\le N_{\rm ROI}$) and that $C_i^{\rm unw}$ is undefined if $k_i = 0$ or $1$.
The global clustering coefficient for the entire network, denoted by $C^{\rm unw}$, is given by the average of $C_i^{\rm unw}$ over all nodes. We exclude the nodes with $k_i\le 1$ from the calculation of $C^{\rm unw}$. Note that $0\le C^{\rm unw}\le 1$. Similar to other types of networks, most brain networks, anatomical or functional, have large values of $C^{\rm unw}$ as compared to randomised networks \cite{Bullmore2009NatRevNeurosci,Bassett2017NatNeurosci}.

\subsection{\label{sub:weighted C}Clustering coefficients for weighted networks}

One can define a weighted functional network by regarding $\rho(i, j)$ as the weight of edge $(i, j)$. Because we do not have established methods to deal with negatively weighted edges (but see \cite{Rubinov2011Neuroimage}) and it is common to discard edges with a negative $\rho(i, j)$ value \cite{Rubinov2010Neuroimage,Kaiser2011Neuroimage}, the weighted adjacency matrix is given by
$w_{ij} = \rho(i, j)$ if $\rho(i, j) > 0$ and $w_{ij}=0$ otherwise. As benchmarks, we consider three variants of weighted clustering coefficient commonly used in the literature \cite{Saramaki2007PhysRevE,Rubinov2010Neuroimage,Rubinov2011Neuroimage,WangGhumare2017NeuralComput}. 
We denote by $(a_{ij})$ the adjacency matrix of the unweighted network obtained by ignoring the edge weight in the weighted network. In other words, we set $a_{ij} = 1$ if $w_{ij}>0$ (equivalently, $\rho(i, j)>0$) and $a_{ij}=0$ otherwise.

The local clustering coefficient of node $i$ proposed by Barrat et al. \cite{Barrat2004PNAS} is given by
\begin{equation}
C_i^{\rm wei,B} = \frac{1}{s_i(k_i-1)} \sum_{\substack{1\le j, \ell\le N_{\rm ROI}\\ j, \ell\neq i}} \frac{w_{ij}+w_{i\ell}}{2} a_{ij} a_{i\ell} a_{j\ell},
\label{eq:C_i^Barrat}
\end{equation}
where $s_i = \sum_{j=1}^{N_{\rm ROI}} w_{ij}$ is the node strength (i.e., weighted degree). It should be noted that $a_{ij} a_{i\ell} a_{j\ell}=1$ if and only if nodes $i$, $j$ and $\ell$ form a triangle in the unweighted network; $a_{ij} a_{i\ell} a_{j\ell}=0$ otherwise. The average of $C_i^{\rm wei,B}$ over all nodes defines the global weighted clustering coefficient denoted by $C^{\rm wei,B}$.

The local clustering coefficient proposed by Onnela et al. \cite{Onnela2005PhysRevE}, which is implemented in the Brain Connectivity Toolbox \cite{Rubinov2010Neuroimage}, is given by
\begin{equation}
C_i^{\rm wei,O} = \frac{1}{k_i(k_i-1)} \sum_{\substack{1\le j,\ell\le N_{\rm ROI}\\ j,\ell\neq i}} \frac{(w_{ij} w_{i\ell} w_{j\ell})^{1/3}}{\max_{i^{\prime}j^{\prime}} w_{i^{\prime}j^{\prime}}}.
\label{eq:C_i^Onnela}
\end{equation}
Factor $\max_{i^{\prime}j^{\prime}}w_{i^{\prime}j^{\prime}}$ normalises $C_i^{\rm wei,O}$ between $0$ and $1$ and prevents it from scaling when the scale of $w_{ij}$ is changed (i.e., when $w_{ij}$ for all $1\le i, j\le N_{\rm ROI}$ is multiplied by the same constant). The corresponding global clustering coefficient, denoted by $C^{\rm wei,O}$, is given by the average of $C_i^{\rm wei,O}$ over all nodes.

The local clustering coefficient proposed by Zhang and Horvath \cite{ZhangHorvath2005StatApplGenetMolBiol} is written as \cite{Saramaki2007PhysRevE}
\begin{equation}
C_i^{\rm wei,Z} = \frac{1}{\max_{i^{\prime}j^{\prime}} w_{i^{\prime}j^{\prime}}}
\frac{\sum_{\substack{1\le j,\ell\le N_{\rm ROI}\\ j,\ell\neq i}} w_{ij}w_{i\ell} w_{j\ell}}
{\sum_{\substack{1\le j,\ell\le N_{\rm ROI}\\ j,\ell\neq i; j\neq \ell}} w_{ij}w_{i\ell}}.
\label{eq:C_i^Zhang}
\end{equation}
The corresponding global clustering coefficient, denoted by $C^{\rm wei,Z}$, is given by the average of $C_i^{\rm wei,Z}$ over all nodes.

\subsection{Our proposal: Clustering coefficients tailored to correlation matrices\label{sub:cor A and M}}

\begin{figure}[t]
\begin{center}
\includegraphics[scale=0.7]{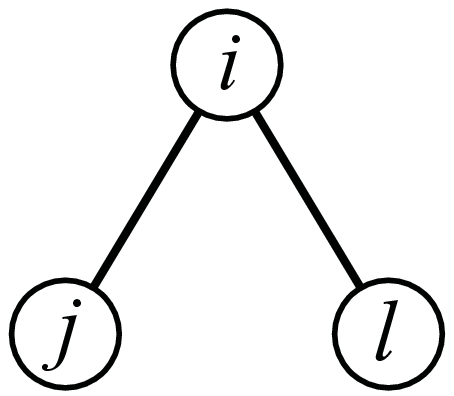}
\caption{Schematic of the indirect path between nodes $j$ and $\ell$ through node $i$.}
\label{fig:schem indirect}
\end{center}
\end{figure}

We propose two clustering coefficient measures for correlation matrices ($C^{\rm cor,A}$ and $C^{\rm cor,M}$). Both of them discount correlation between ROIs $j$ and $\ell$ that is expected from the correlation between ROIs $i$ and $j$ and that between $i$ and $\ell$, i.e., indirect path between $j$ and $\ell$ through $i$ (Fig.~\ref{fig:schem indirect}) \cite{Zalesky2012Neuroimage}. One measure uses the three-way partial correlation coefficient and the other measure uses the partial mutual information.

The three-way partial correlation coefficient between ROIs $j$ and $\ell$ controlling for the influence of ROI $i$, denoted by ${\rho}^{\text{partial}}(j, \ell \mid i)$, is defined by \cite{Whittaker1990book}
\begin{equation}
\rho^{\rm partial}(j, \ell \mid i) = \frac{\rho(j,\ell) - \rho(i,j) \rho(i,\ell)}
{\sqrt{1-\rho^2(i,j)} \sqrt{1-\rho^2(i,\ell)}}.
\label{eq:def partial cor}
\end{equation}
Equation~\eqref{eq:def partial cor} indicates that ROIs $i$ and $j$ would be correlated with an amount
$\rho(i,j) \rho(i,\ell)$ by default owing to the indirect path between $j$ and $\ell$ through $i$ (e.g., \cite{Zalesky2012Neuroimage}). Deviations of $\rho(j,\ell)$ from $\rho(i,j) \rho(i,\ell)$ quantify the tendency that $j$ and $\ell$ are more strongly or weakly connected than is expected from the presence of an indirect path between $j$ and $\ell$ through $i$. Based on this observation, we define a first variant of the clustering coefficient as follows.

It is difficult to interpret negative correlation values in functional connectivity data \cite{Fox2009JNeurophysiol,Murphy2009Neuroimage,Rubinov2011Neuroimage,SmithMiller2011Neuroimage,Sporns2011book,Fornito2013Neuroimage}. Therefore, we assume that any deviation of $\rho(j,\ell)$ from $\rho(i,j)\rho(i,\ell)$ caused by the effect of $i$, irrespective of whether it is positive or negative, contributes to the local clustering coefficient at $i$. We define the local clustering coefficient for ROI $i$, denoted by $C_i^{\rm cor,A}$ (superscript A standing for the absolute value), as
\begin{equation}
C_i^{\rm cor,A} = \frac{\sum_{\substack{1\le j< \ell\le N_{\rm ROI}\\ j,\ell\neq i}} \left|\rho(i,j) \rho(i,\ell) \rho^{\rm partial}(j,\ell \mid i)\right|}
{\sum_{\substack{1\le j< \ell\le N_{\rm ROI}\\ j,\ell\neq i}} \left|\rho(i,j) \rho(i,\ell)\right|}.
\label{eq:C_i^A}
\end{equation}
In other words, $C_i^{\rm cor,A}$ is a weighted average of the absolute value of the partial correlation over pairs of $j$ and $\ell$. We have employed the weight $\left|\rho(i,j) \rho(i,\ell)\right|$ for averaging because a high clustering around ROI $i$ should imply strong association between ROIs $j$ and $\ell$ (in the sense of partial correlation) when $i$ and $j$ are strongly connected and $i$ and $\ell$ are. We have used $\rho^{\rm partial}(j,\ell \mid i)$ instead of $\overline{\rho}^{\rm partial}(j, \ell)$, i.e., the partial correlation between $j$ and $\ell$ controlling for the effect of the other $N_{\rm ROI}-2$ ROIs, to make $C_i^{\rm cor,A}$ a locally calculated quantity as is the case for the clustering coefficients for networks (e.g., $C_i^{\rm und}$, $C_i^{\rm wei, B}$, $C_i^{\rm wei, O}$ and $C_i^{\rm wei, Z}$). The corresponding global clustering coefficient, denoted by $C^{\rm cor,A}$, is given by the average of $C_i^{\rm cor,A}$ over all nodes. Note that $0\le C_i^{\rm cor,A}\le 1$ $(1\le i\le N_{\rm ROI})$ and $0\le C^{\rm cor,A}\le 1$.

We also use another definition of the clustering coefficient based on the partial mutual information, which is a nonlinear correlation measure \cite{Frenzel2007PhysRevLett}. By definition, the mutual information is nonnegative and invariant under flipping of the sign of the random variable. We use the partial mutual information between ROIs $j$ and $\ell$ conditioned on ROI $i$ in place of $\rho^{\rm partial}(j,\ell \mid i)$ to define the second variant of the local clustering coefficient for correlation matrices, denoted by $C^{\rm cor,M}_i$ (superscript M standing for the mutual information).

The partial mutual information is defined as
\begin{equation}
I(X_j, X_{\ell} \mid X_i) = h(X_j,X_i) + h(X_{\ell},X_i) - h(X_i) - h(X_j,X_{\ell},X_i),
\label{eq:partial mutual info}
\end{equation}
where $X_i$, $X_j$ and $X_{\ell}$ are the random variables on ROIs $i$, $j$ and $\ell$, respectively, and $h$ is the (joint) entropy. For example, $h(X_i) = - \sum_x p(x)\log_2 p(x)$, where $p(x)$ is the probability that $X_i = x$, and $h(X_j, X_i) = - \sum_{x,x^{\prime}} p(x,x^{\prime}) \log_2 p(x,x^{\prime})$, where $p(x,x^{\prime})$ is the probability that $(X_j, X_i) = (x,x^{\prime})$. By assuming that the BOLD signals at ROIs $i$, $j$ and $\ell$ obey a multivariate Gaussian distribution, one obtains the entropy values in Eq.~\eqref{eq:partial mutual info} as follows \cite{Rieke1997-1999book,CoverThomas2006book,Frenzel2007PhysRevLett}:
\begin{equation}
h(X_{\alpha_1}, \ldots, X_{\alpha_d}) = \frac{d}{2}(1+\ln 2\pi) + \frac{1}{2}\ln \det \text{cov}^{\prime},
\label{eq:joint h for Gaussian}
\end{equation}
where $d$ is the number of random variables and $\text{cov}^{\prime} = (\text{cov}^{\prime}_{ij})$ is the $d\times d$ covariance matrix constructed by $X_{\alpha_1}$, $\ldots$, $X_{\alpha_d}$, i.e., $\text{cov}^{\prime}_{ij} = \text{E}\left[ X_{\alpha_i} X_{\alpha_j}\right]$, where $\text{E}\left[\cdot\right]$ represents the expectation. By substituting Eq.~\eqref{eq:joint h for Gaussian} in Eq.~\eqref{eq:partial mutual info} and setting $\text{cov}^{\prime}_{ij} = \rho(i,j)$, we obtain
\begin{align}
I(X_j, X_{\ell} \mid X_i) =& \frac{1}{2}\left[ 
\ln\left(1-\rho^2(i,j)\right) + \ln\left(1-\rho^2(i,\ell)\right) \right.\notag\\
-& \left. \ln\left(1-\rho^2(i,j)-\rho^2(i,\ell)-\rho^2(j,\ell) + 2\rho(i,j)\rho(i,\ell)\rho(j,\ell)\right)\right].
\label{eq:partial mutual info 2}
\end{align}
Using the partial mutual information, we define
\begin{equation}
C_i^{\rm cor,M} = \frac{\sum_{\substack{1\le j< \ell\le N_{\rm ROI}\\ j,\ell\neq i}} \left|\rho(i,j) \rho(i,\ell) \right|
I(X_j, X_{\ell} \mid X_i)}
{\frac{1+\ln 2\pi}{2} \sum_{\substack{1\le j< \ell\le N_{\rm ROI}\\ j,\ell\neq i}} \left|\rho(i,j) \rho(i,\ell)\right|}.
\label{eq:C_i^M}
\end{equation}
The denominator normalises the $C_i^{\rm cor,M}$ value to range between $0$ and $1$. The corresponding global clustering coefficient, denoted by $C^{\rm cor,M}$, is given by the average of $C_i^{\rm cor,M}$ over all nodes.

As a robustness test, we also examined variants of these clustering coefficients constrained to only positive triangles or negative triangles. We define  $C^{{\rm cor, A}, +}$ by restricting the enumeration of triangles in the calculation of $C^{\rm cor, A}$ to the positive triangles. In other words, we restrict the summation on the numerator and denominator of Eq.~\eqref{eq:C_i^A} to $j$ and $\ell$ satisfying $\rho(i,j)$, $\rho(i,\ell)$, $\rho(j,\ell) > 0$. We similarly define $C^{{\rm cor,A},-}$, $C^{{\rm cor,M},+}$ and $C^{{\rm cor,M},-}$. We removed six participants from the calculation of $C^{{\rm cor,A},-}$ and $C^{{\rm cor,M},-}$. This is because, for these participants, there was at least one ROI $i$ at which there was no triangle with $\rho(i,j)$, $\rho(i,\ell)$, $\rho(j,\ell) < 0$, rendering $C^{{\rm cor,A},-}$ and $C^{{\rm cor,M},-}$ undefined.

We provided C++ code for calculating the proposed clustering coefficients on Github
(https://github.com/naokimas/clustering-corr-mat).

\subsection{H-Q-S algorithm}

As a null model of the covariance matrix, we employed the Hirschberger-Qu-Steuer (H-Q-S) algorithm \cite{Hirschberger2007EurJOperRes}. As recent fMRI data analysis has demonstrated, the H-Q-S algorithm is a more suitable null model than conventional null models in which the topology is randomised
\cite{Zalesky2012Neuroimage,Hosseini2013PlosOne}. The H-Q-S algorithm preserves the mean of the diagonal elements, the mean of the off-diagonal elements and the variance of the off-diagonal elements of the given covariance matrix. From the fMRI data of each participant, we obtained the covariance matrix in the course of calculating the functional connectivity, which is the correlation matrix. Based on this covariance matrix, we generated random covariance matrices using H-Q-S algorithm. We then converted the generated random covariance matrices into correlation matrices, which were used as randomised functional connectivity matrices. We did not implement a fine-tuned heuristic variant proposed in \cite{Zalesky2012Neuroimage}.

Denote by $\mu_{\rm on}$ the average of the diagonal elements of the covariance matrix over the $N_{\rm ROI}$ diagonal elements. Denote by $\mu_{\rm off}$ and $\sigma_{\rm off}^2$ the average and variance of the off-diagonal elements, respectively. We set $\overline{t}_{\max}= \max \left( 2, \lfloor \left(\mu_{\rm on}^2 - \mu_{\rm off}^2\right)/\sigma_{\rm off}^2\rfloor\right)$, where $\lfloor \cdot\rfloor$ is the largest integer smaller than or equal to the argument. Then, we generate $N_{\rm ROI}\times \overline{t}_{\max}$ variables, denoted by $x_{i,t}$ ($1\le i\le N_{\rm ROI}$, $1\le t\le \overline{t}_{\max}$) that independently obey the normal distribution with mean $\sqrt{\mu_{\rm off}/\overline{t}_{\max}}$ and variance $-\mu_{\rm off}/\overline{t}_{\max} +
\sqrt{\mu_{\rm off}^2/\overline{t}_{\max}^2 + \sigma_{\rm off}^2/\overline{t}_{\max}}$. The H-Q-S algorithm generates a randomised covariance matrix by $\text{cov}_{ij} = \sum_{t=1}^{\overline{t}_{\max}} x_{i\ell} x_{j\ell}$ ($1\le i, j\le N_{\rm ROI})$. In other words, the algorithm assumes that the signal at ROI $i$ is a white-noise time series with a positive bias of length $\overline{t}_{\max}$, which is independent across the time and ROIs.

\subsection{White-noise signals}

To generate another null model of the covariance matrix, we used white-noise signals. For each ROI, we generated a time series of length 200 in which the signal at each time step and ROI independently obeyed the normal distribution with mean 0 and standard deviation 1. Then, we calculated the covariance matrix using pairs of the $N_{\rm ROI}$ time series and converted it into the correlation matrix.

\subsection{Participants}

One-hundred thirty eight ($n=138$) healthy and right-handed participants (54 females and 84 males) were selected from the Nathan Kline Institute's (NKI) Rockland Sample \cite{Nooner2012FrontNeurosci}.
The NKI's data that we used are publicly available.
%
% (http://fcon\_1000.projects.nitrc.org/indi/pro/nki.html). 
%
The age of the participants ranged between 18 and 85 years (mean $= 41.7$,  std $= 18.4$). 

For four of our participants, the H-Q-S algorithm did not work because the average off-diagonal element for the empirical covariance matrix was negative, violating the precondition for the algorithm \cite{Hirschberger2007EurJOperRes}. Therefore, we removed the four participants in the analysis that used the H-Q-S algorithm.

\subsection{fMRI data acquisition and preprocessing}

The MRI data were recorded in a 3T scanner (MAGNETOM, TrioTim syngo MR B15, Siemens). fMRI data were obtained during rest with an echo planner imaging (EPI) sequence (TR $=$ 2500 ms, TE $=$ 30 ms, flip angle $= 80^\circ$, 38 slices, spatial resolution $= 3\times 3\times 3$ mm$^3$, FOV $=$ 216 ms, acquisition time $=$ 10 m 55 s). A total of $t_{\max}=258$ volumes was recorded from each participant. Anatomical images were acquired with T1-weighted sequence (MPRAGE) (TR = 2500ms, TE $=$ 3.5 ms, flip angle $= 8^\circ$, spatial resolution $= 1\times 1\times 1$ mm$^3$). During the EPI data acquisition, the participants were asked to be relaxed with their eyes open. 

Data preprocessing was performed using FMRIB's Software Library (FSL; www.fmrib.ox.ac.uk/fsl), including skull stripping of structural images with BET and registration with FLIRT; each functional image was registered to the participant's high-resolution brain-extracted structural image and the standard Montreal Neurological Institute (MNI) 2-mm brain. Functional data were then preprocessed with motion correction with MCFLIRT and smoothing with full-width half-maximum 5 mm. We also applied additional preprocessing steps to the functional data to remove spurious variance. First, we regressed out six head motion parameters, the global signal, cerebrospinal fluid (CSF) signal, and white matter (WM) signal with FSL FEAT. For each participant, CSF, gray matter (GM) and WM were segmented through FSL's FAST based on his/her T1. The signal averaged over all voxels in GM, WM and CSF was used as global signal. We then applied band-pass temporal filtering (0.01--0.1 Hz).

%\subsection{\del{Executive score}}

% \del{We constructed the executive score of the participants in the same manner as in our previous study [Ezaki2017submitted]. For completeness, we replicate how we did it. Participants' executive functions were assessed by the Delis-Kaplan Executive Function System (D-KEFS) [Delis2001book,Delis2004JIntNeuropsycholSoc] which consists of multiple cognitive tests to assess executive functions. In accordance with previous literature [Latzman2010Assessment,Barbey2012Brain], we constructed a composite executive score for each individual by applying a factor analysis to scores from five tests in D-KEFS: verbal fluency, sorting task, 20 questions, color word task, and design fluency. The analysis revealed one factor with the eigenvalue greater than 1 (the engenvalue for this factor was 2.18); therefore, we used the factor score for this first factor as an executive function score. Participants' IQ was assessed by the Wechsler Abbreviated Scale of Intelligence (WASI) [Wechsler1999book]; a full scale intelligence quotient, verbal IQ, and performance IQ scores were used in the current study.}

\subsection{Linear mixed model}

To estimate the linear mixed model with a fixed effect and random effects, we used the \textit{lmer} function in lme4 package in R (v.\;3.4.1). The dependent variable in the linear mixed model was the local clustering coefficient. The fixed and random effects were the node strength and the participant, respectively.
To obtain the $P$ value, we used the \textit{F}-test with Kenward-Roger approximation implemented as the \textit{KRmodcomp} function in pbkrtest package in R.

\section{Results}

We demonstrate the utility of the proposed clustering coefficients on fMRI data collected from participants of a wide range of the age. We looked for associations of the clustering coefficients with the age and its dependence on the ROIs.

\subsection{Comparison with null models}

Statistically larger values of conventional clustering coefficients have repeatedly been observed in empirical brain networks as compared to the null models \cite{Bullmore2009NatRevNeurosci,Bassett2017NatNeurosci}.
Motivated by these studies, we examined whether the amount of clustering was different between the empirical data and these null models after we controlled for the amount of correlation between two ROIs $j$ and $\ell$ expected from an indirect path between $j$ and $\ell$ through a third ROI $i$. For each participant, we compared the proposed clustering coefficients between the fMRI data obtained from all the participants, those calculated for the H-Q-S null model \cite{Hirschberger2007EurJOperRes,Zalesky2012Neuroimage}, and white-noise signals. 

The empirical correlation matrices yielded significantly larger values of the clustering coefficient than the correlation matrices for white-noise signals did. The results were consistent between the two definitions of the clustering coefficient, i.e., $C^{\rm cor,A}$ (empirical: $0.221 \pm 0.029$, white noise: $0.057 \pm 0.002$, $t_{137}=66.0$, $P<10^{-6}$, $d=11.28$) and
$C^{\rm cor,M}$ (empirical: $0.031 \pm 0.008$, white noise: $0.002 \pm 0.000$, $t_{137}=40.3$, $P<10^{-6}$, $d=6.89$). This result is consistent with the previous findings with the conventional clustering coefficients for networks, where empirical functional networks tended to have large clustering coefficients than randomised networks \cite{Salvador2005CerebCortex,Eguiluz2005PhysRevLett,Achard2006JNeurosci,Bassett2006Neuroscientist}.

In contrast, the two types of clustering coefficient were smaller for the empirical data than for the randomised data generated by the H-Q-S algorithm (for $C^{\rm cor,A}$, H-Q-S: $0.281 \pm 0.073$, $t_{133}=-12.4$, $P<10^{-6}$, $d=-2.15$; for $C^{\rm cor,M}$, H-Q-S: $0.056 \pm 0.039$, $t_{133}=-8.59$, $P<10^{-6}$, $d=-1.49$). This result has probably arisen because the H-Q-S algorithm generates a correlation matrix from short white-noise time series assumed at each ROI. Then, the partial correlation (Eq.~\eqref{eq:def partial cor}) calculated for the H-Q-S algorithm is distributed relatively widely due to statistical fluctuations, whose distribution can be even wider than that for the empirical data. This fact makes $C^{\rm cor, A}$ and $C^{\rm cor,M}$, which more or less depends on the absolute value of the partial correlation, large for the randomised data generated by the H-Q-S algorithm.

\subsection{Age-related differences in the clustering coefficients tailored to correlation matrices\label{sub:corr with age and exec}}

\begin{figure}[t]
\begin{center}
\includegraphics[scale=0.25]{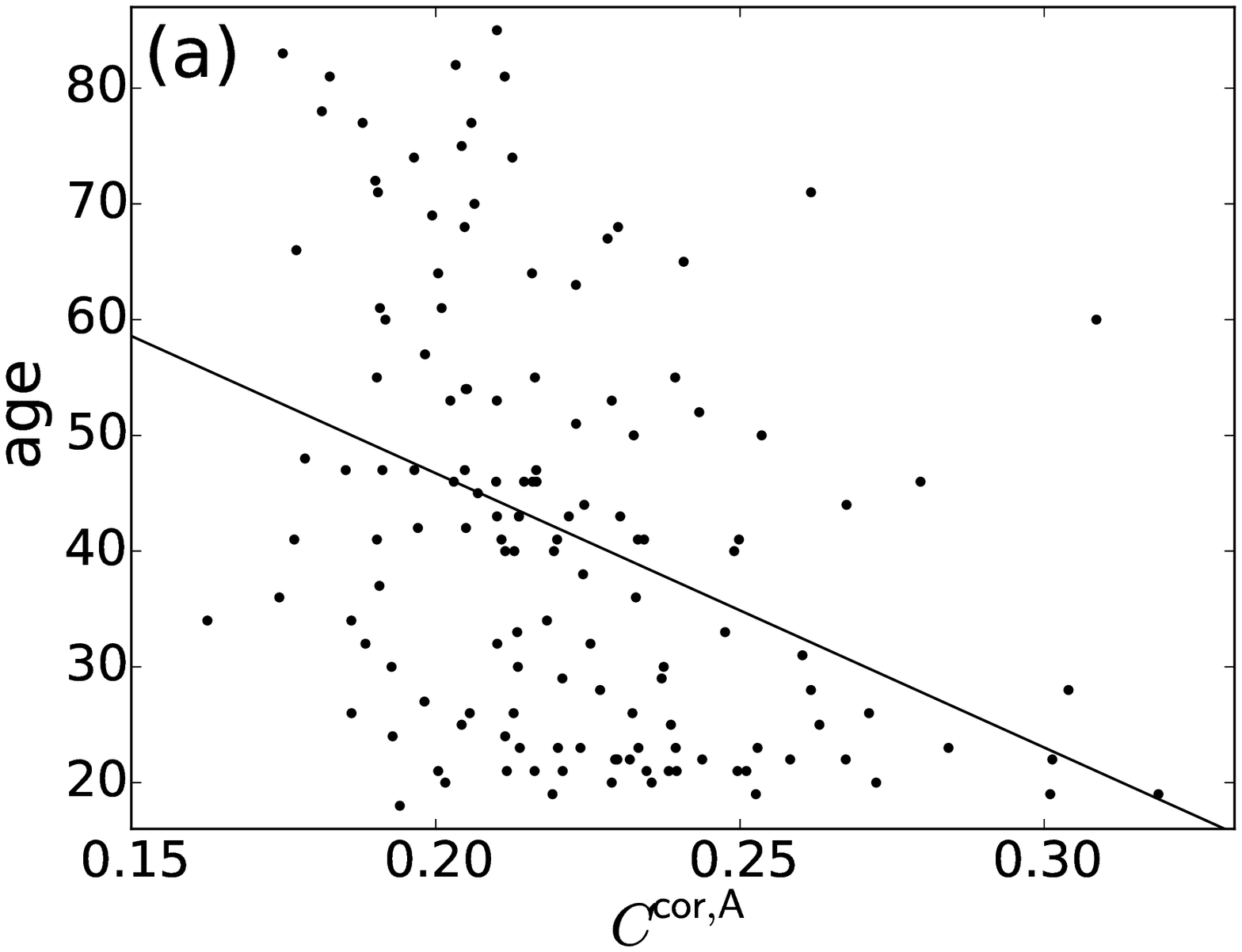}
\includegraphics[scale=0.25]{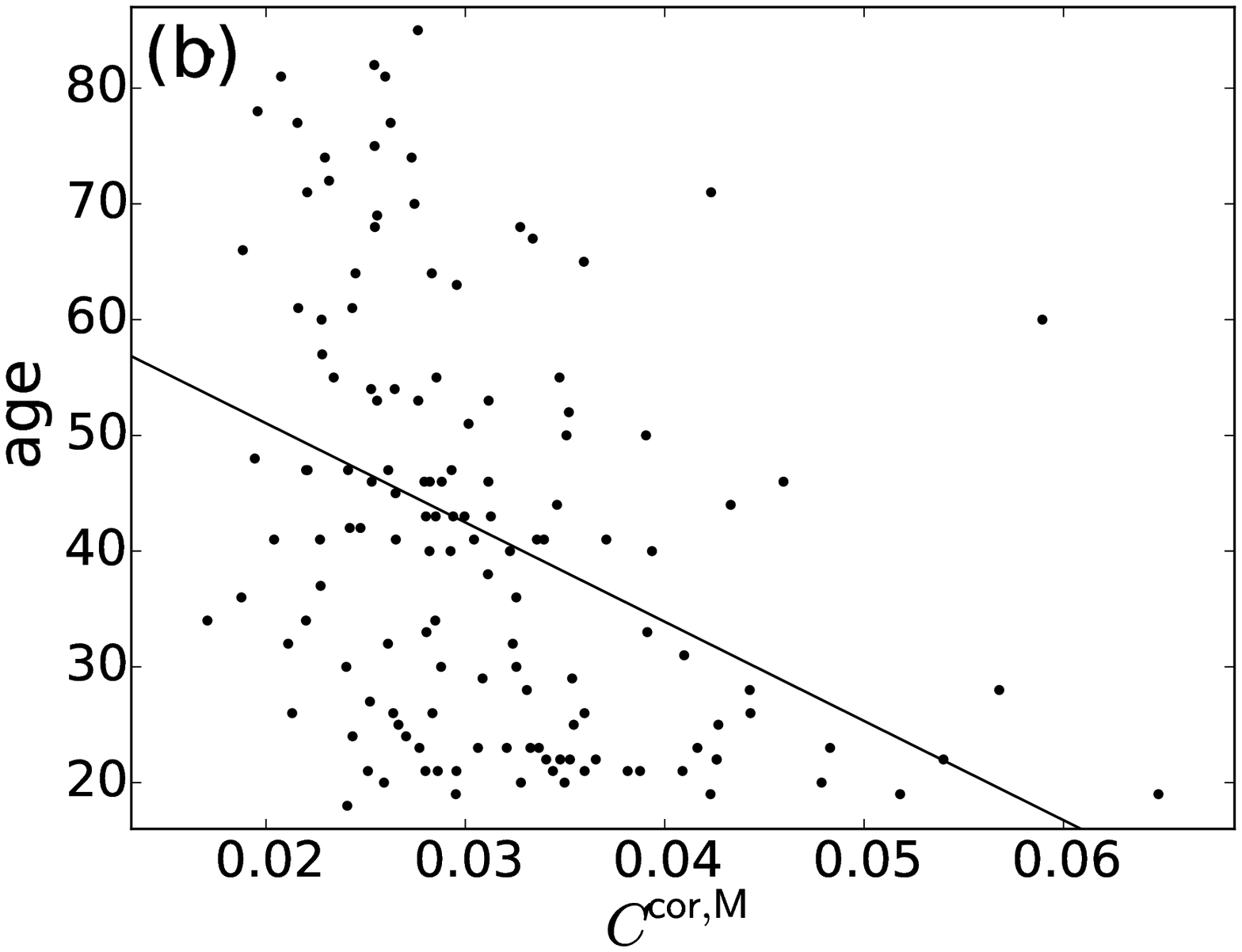}
\includegraphics[scale=0.25]{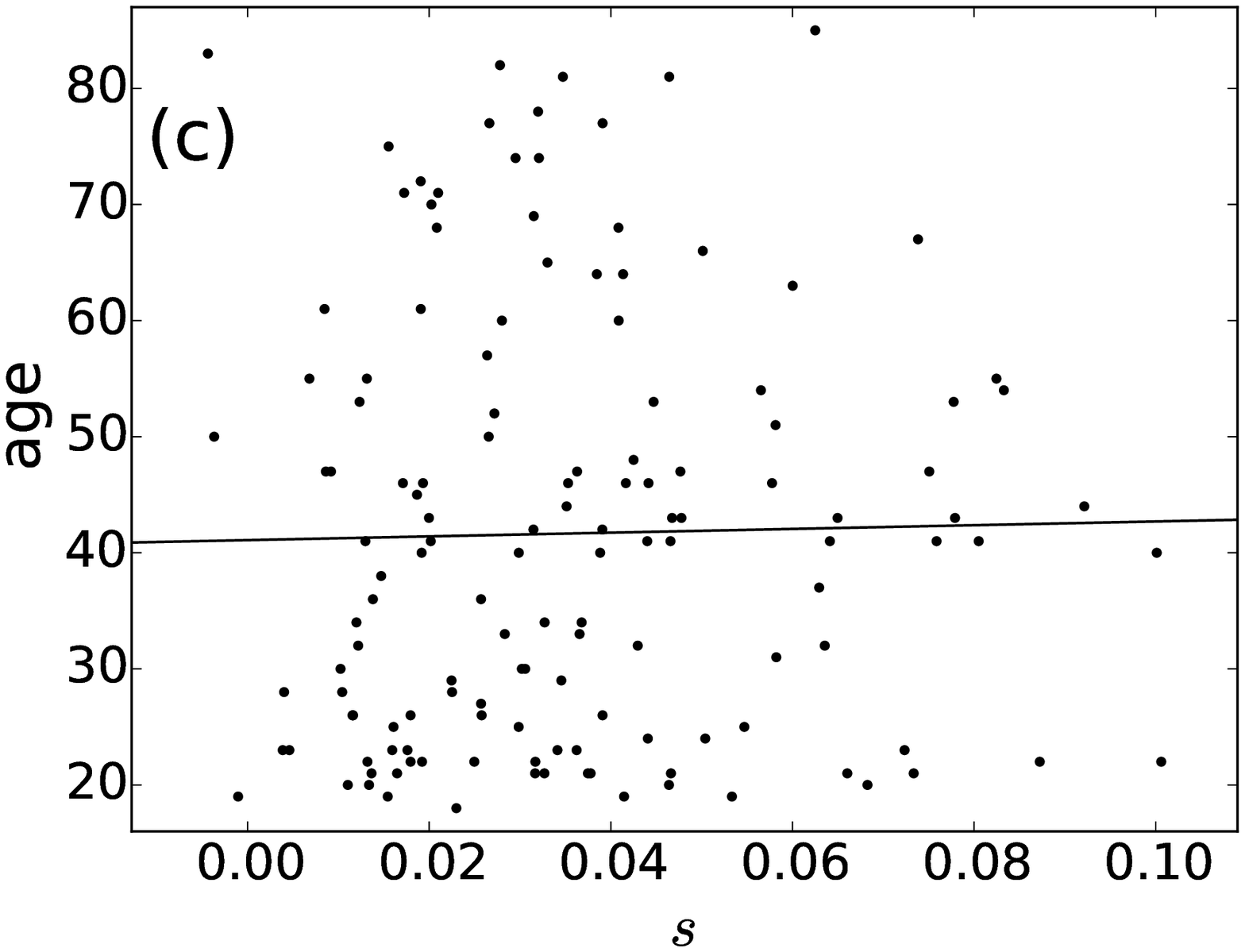}
\includegraphics[scale=0.25]{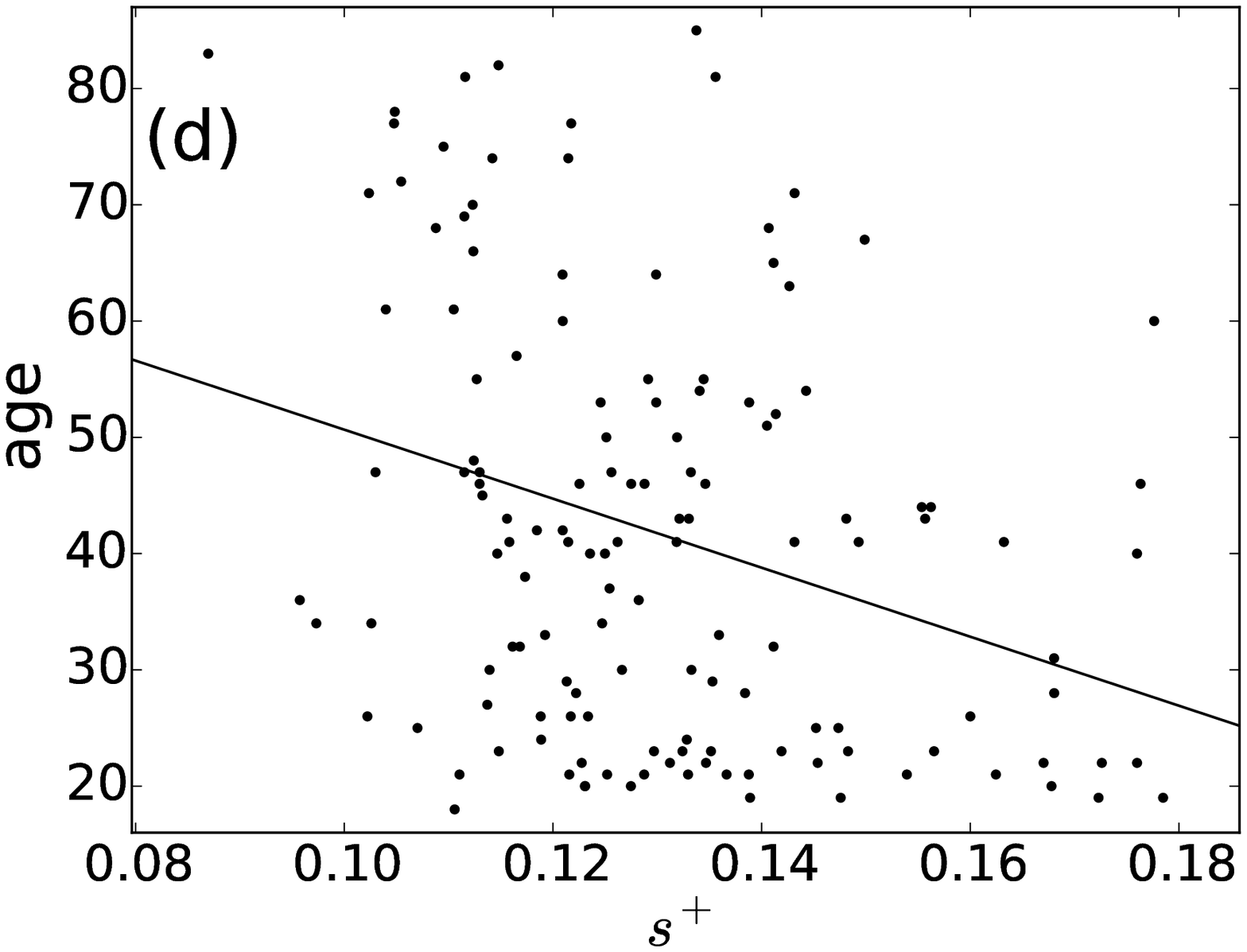}
\includegraphics[scale=0.25]{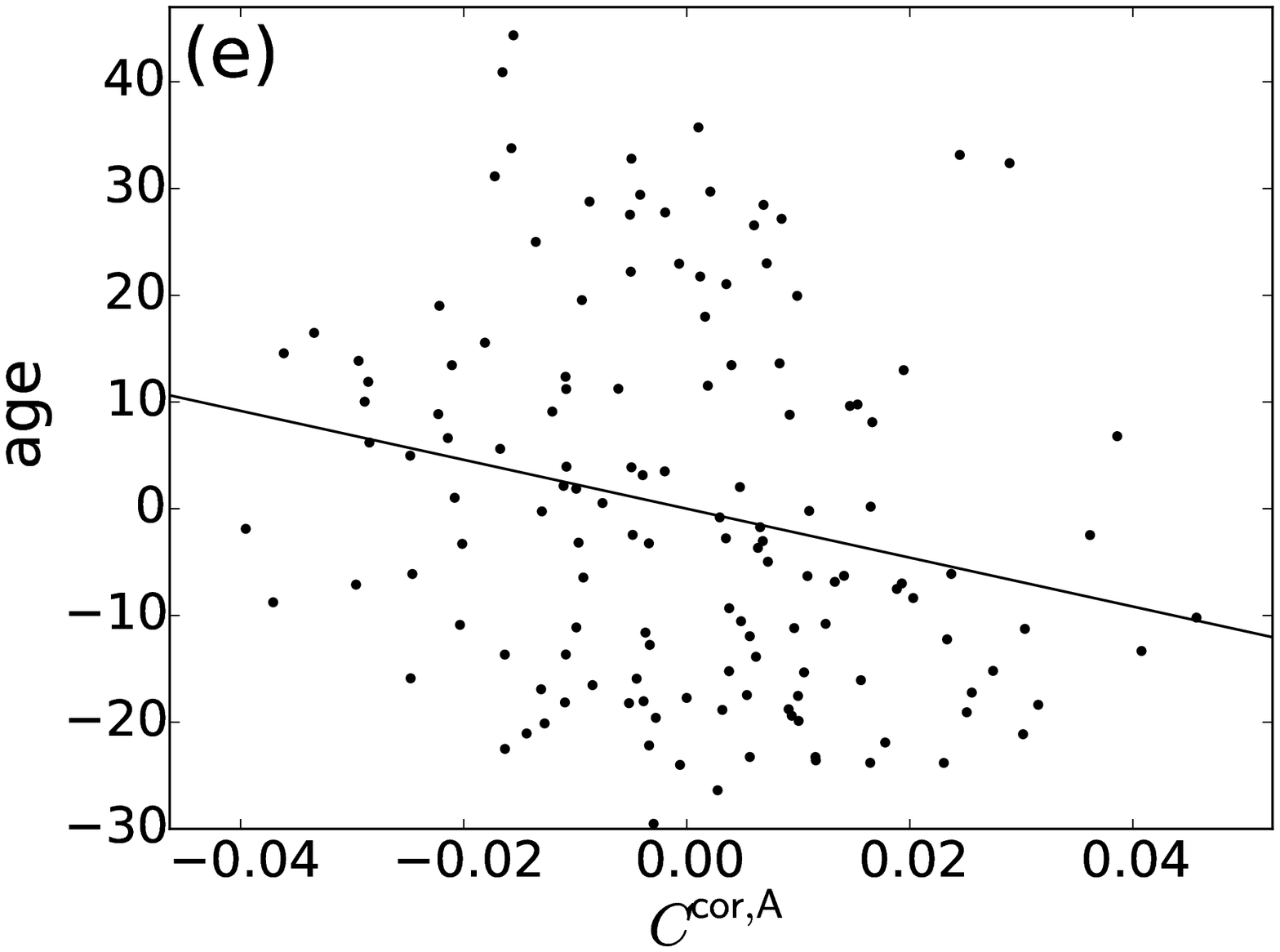}
\includegraphics[scale=0.25]{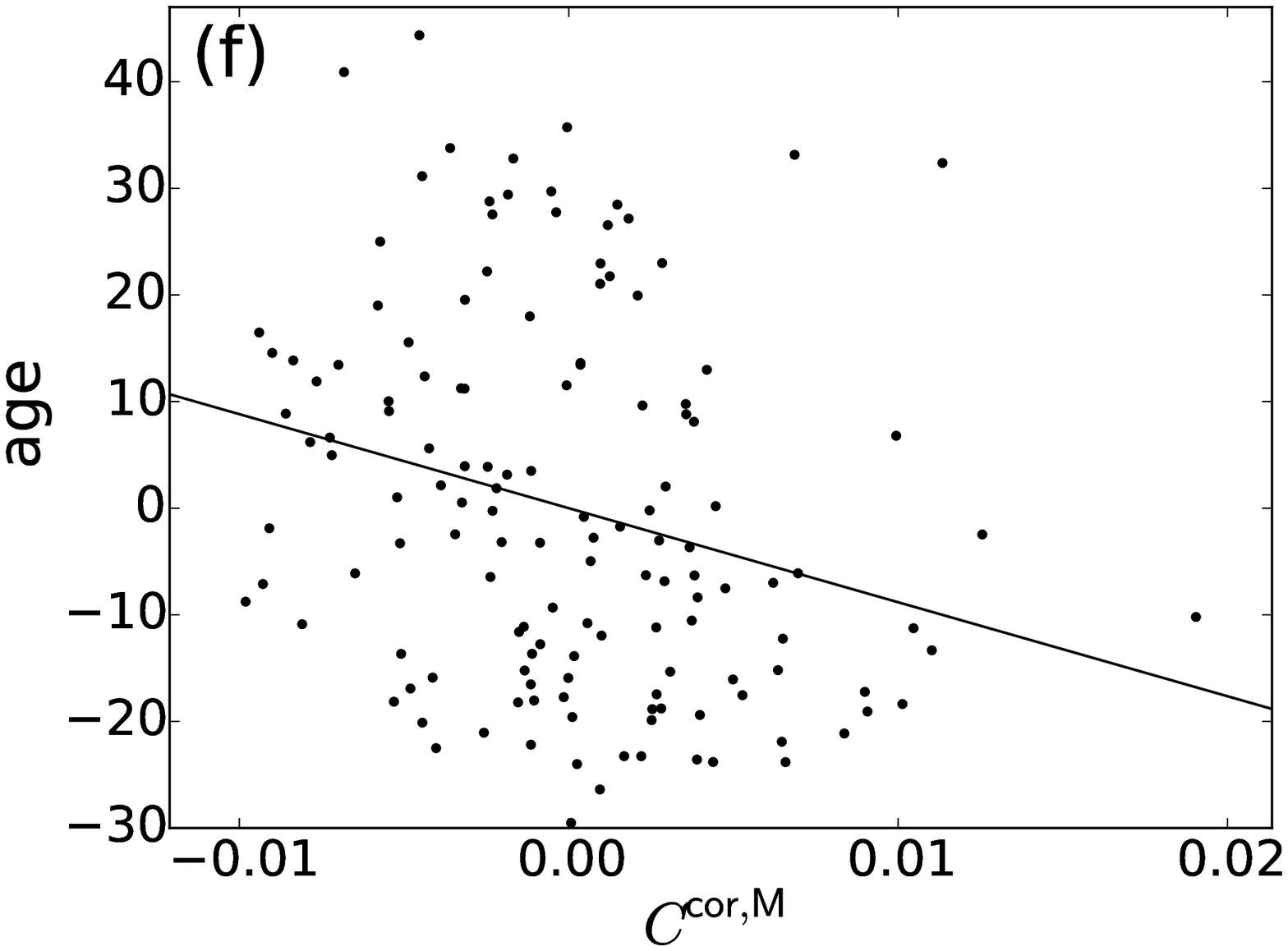}
\caption{Relationship between the age and network indices. (a) $C^{\rm cor,A}$ vs age. (b) $C^{\rm cor,M}$ vs age. (c) $s$ vs age. (d) $s^+$ vs age. (e) $C^{\rm cor,A}$ vs age, where the effect of $s^+$ is regressed out. (f) $C^{\rm cor,M}$ vs age, where the effect of $s^+$ is regressed out. A symbol represents an individual. The lines represent the linear fit: (a) age $= -237.0 \times C^{\rm cor,A} + 94.1$, (b) age $= -857.5 \times C^{\rm cor,M} + 68.2$, (c) age $= 16.1 \times s + 41.1$, (d) age $= -296.8 \times s^+ + 80.3$, (e) age $= -229.2 \times C^{\rm cor,A}$, (f) age $= -882.0 \times C^{\rm cor,M}$. In (e) and (f), the linear contribution of $s^+$ to the variables plotted in (a) and (b) are subtracted from the original variables and the residuals are plotted. The Pearson correlation coefficient between the residuals gives the partial correlation coefficient.}
\label{fig:corr with age}
\end{center}
\end{figure}

Normal ageing was shown to adversely affect small-worldness of brain networks \cite{Achard2007PlosComputBiol}. Because the clustering coefficient is a major index which is used to assess the small-worldness of networks \cite{Watts1998Nature}, we examined whether our clustering coefficients were able to detect such age-related changes in network structure. We found a negative relationship between each of the two types of clustering coefficients (i.e., $C^{\rm cor,A}$ and $C^{\rm cor,M}$) and the age ($C^{\rm cor,A}$: $r_{136} = -0.377$, $P<10^{-5}$; $C^{\rm cor,M}$: $r_{136}=-0.397$, $P<10^{-5}$; Fig.~\ref{fig:corr with age}(a), (b), Table~\ref{tab:corr with age}).
To explore whether the age is correlated with an index that can be more easily calculated than the clustering coefficient, we examined the relationships between the age and two indices of average functional connectivity. We found that the age was uncorrelated with $s$ ($r_{136} = 0.020$, $P=0.82$; Fig.~\ref{fig:corr with age}(c), Table~\ref{tab:corr with age}) but negatively correlated with $s^+$ ($r_{136} = -0.311$, $P=0.0002$; Fig.~\ref{fig:corr with age}(d), Table~\ref{tab:corr with age}). The two clustering coefficients were also strongly correlated with $s^+$, whereas they were not correlated with $s$ (Table~\ref{tab:corr with node strength}). Therefore, we suspected that the negative correlation between the clustering coefficients and the age was caused by the combination of the negative correlation between $s^+$ and the age and the positive correlation between $s^+$ and the clustering coefficient. However,
significant negative correlation persisted between the clustering coefficients and the age even after controlling for the effect of $s^+$ ($C^{\rm cor,A}$: $r_{136} = -0.224$, $P=0.0076$; $C^{\rm cor,M}$: $r_{136} = -0.259$, $P=0.0019$; see Fig.~\ref{fig:corr with age}(e) and (f) for the scatter plot between the clustering coefficient and the age after the linear effect of $s^+$ has been regressed out from both variables; also see Table~\ref{tab:corr with age}). This result indicates that the negative correlation between the clustering coefficients and age is not completely explained by $s^+$. Therefore, $C^{\rm cor, A}$ and $C^{\rm cor, M}$ quantify effects of the age on fMRI signals beyond what is revealed by the average functional connectivity.

% \del{For another proof of concept, we also examined the relationship between the proposed clustering coefficients (i.e., $C^{\rm cor, A}$ and $C^{\rm cor, M}$) and the participant's executive score. First of all, the executive score was negatively correlated with the age ($r_{131} = -0.405$, $P = 1.3\times 10^{-6}$), which is consistent with the previous results [Park2002PsycholAging,Salthouse2009JIntNeuropsycholSoc,Ezaki2017submitted]. We found a positive correlation between the clustering coefficients and the executive score (Table~\ref{tab:corr with exec}) although the correlation was smaller than with the age. These results are consistent with those obtained for the correlation with the age (Fig.~\ref{fig:corr with age}(a), (b), Table~\ref{tab:corr with age}), because high-performing participants, who tend to be younger adults, would have a large clustering coefficient value. When we controlled for the effect of $s^+$, we found a mildly positive correlation between the clustering coefficient and the executive score (Table~\ref{tab:corr with exec}).}

Positive edges and negative edges may have distinct meanings \cite{Rubinov2011Neuroimage}. Therefore, we examined variants of the proposed clustering coefficients calculated only from positive triangles (denoted by
$C^{{\rm cor,A},+}$ and $C^{{\rm cor,M},+}$) or negative triangles
(denoted by $C^{{\rm cor,A},-}$ and $C^{{\rm cor,M},-}$). 
These variants of clustering coefficients were negatively correlated with the age
 ($C^{{\rm cor,A},+}$: $r_{136} = -0.398$, $P<10^{-5}$;
 $C^{{\rm cor,A},-}$: $r_{130} = -0.291$, $P = 0.0007$;
 $C^{{\rm cor,M},+}$: $r_{136}=-0.431$, $P<10^{-5}$;
 $C^{{\rm cor,M},-}$: $r_{130}=-0.304$, $P = 0.0004$).
This negative relationship was significant even after controlling for the effect of $s^+$
($C^{{\rm cor,A},+}$: $r_{136} = -0.263$, $P = 0.0019$;
$C^{{\rm cor,A},-}$: $r_{130} = -0.197$, $P = 0.024$;
$C^{{\rm cor,M},+}$: $r_{136} = -0.315$, $P = 0.0002$;
$C^{{\rm cor,M},-}$: $r_{130} = -0.196$, $P = 0.024$). The negative correlation was stronger for the clustering coefficients based on the positive triangles
(i.e., $C^{{\rm cor,A},+}$ and $C^{{\rm cor,M},+}$) than those based on the negative triangles
(i.e., $C^{{\rm cor,A},-}$ and $C^{{\rm cor,M},-}$). We conclude that the age-related differences in the clustering coefficients observed with $C^{\rm cor,A}$ and $C^{\rm cor,M}$ are robust against the restriction of the method to the positive or negative triangles.
Note that the age-related decline of $C^{{\rm cor,A},+}$ and $C^{{\rm cor,M},+}$ was stronger than that of $C^{\rm cor,A}$ and $C^{\rm cor,M}$, respectively.

The rationale behind our clustering coefficients is that the correlation between two neighbours of a focal ROI should be discounted due to the effect of the indirect path. The clustering coefficients $C^{\rm cor, A}$ and $C^{\rm cor, M}$ are not the only indices complying with this rationale. To examine the robustness of our results with respect to specific definitions of the clustering coefficient, we examined the relationship among two other variants of the clustering coefficient designed for correlation matrices and $s$, $s^+$ and the age. Although the correlation between the clustering coefficient and the age was somewhat weaker than in the case of $C^{\rm cor, A}$ and $C^{\rm cor, M}$, the results with the other two variants of the clustering coefficient were qualitatively the same as those for $C^{\rm cor, A}$ and $C^{\rm cor, M}$ (Appendix~A). 

\subsection{Age-related differences in the conventional clustering coefficients}

We repeated the same analysis using the clustering coefficients previously proposed for unweighted networks (i.e., $C^{\rm unw}$) and weighted networks (i.e., $C^{\rm wei,B}$, $C^{\rm wei,O}$ and $C^{\rm wei,Z}$). For unweighted networks, we used two edge density values, 0.1 and 0.2.
Qualitatively, the clustering coefficients for unweighted and weighted networks behaved similarly to $C^{\rm cor,A}$ and $C^{\rm cor,M}$ did. In other words, the clustering coefficients were negatively correlated with the age (Table~\ref{tab:corr with age}), positively and strongly correlated with $s^+$ and not with $s$ with the exception of $C^{\rm wei, B}$  (Table~\ref{tab:corr with node strength}). However, the correlation with the age was weaker than in the case of $C^{\rm cor,A}$ and $C^{\rm cor,M}$ (Table~\ref{tab:corr with age}; see Appendix~B for the statistical results). In fact, the partial correlation between the conventional clustering coefficients (i.e., $C^{\rm unw}$, $C^{\rm wei,B}$, $C^{\rm wei,O}$ and $C^{\rm wei,Z}$) and the age was not significant when one controls the effect of $s^+$ (Table~\ref{tab:corr with age}).
These results suggest that these conventional clustering coefficients extract relatively similar information to that extracted by $s^+$ as compared to $C^{\rm cor,A}$ and $C^{\rm cor,M}$ do.

\subsection{Age-related differences in the clustering coefficients for networks derived from partial correlation matrix}

Functional networks are often defined in terms of the partial correlation matrix \cite{Salvador2005CerebCortex,Marrelec2006Neuroimage,SmithMiller2011Neuroimage}. Therefore, as a benchmark, we calculated the conventional clustering coefficients (for unweighted and weighted networks) 
for functional networks defined by the partial correlation matrix. The clustering coefficients were not correlated with $s$ or $s^+$ (Table~\ref{tab:corr with node strength}). These clustering coefficients were also uncorrelated with the age (Table~\ref{tab:corr with age}).

\subsection{Relationship between the local clustering coefficients and the node strength (weighted degree of the node)\label{sub:local C}}

Local clustering coefficients have been used for characterising individual ROIs \cite{SpornsZwi2004Neuroinfo,Achard2006JNeurosci,He2007CerebCortex,Alexanderbloch2010FrontSystNeurosci,Lynall2010JNeurosci,Power2010Neuron,Vandenheuvel2010JNeurosci,Vandenheuvel2011JNeurosci,Wee2011Neuroimage,Fornito2012Neuroimage,Tijms2013NeurobiolAging,Salallonch2014NeurobiolAging}. In this section we show that, differently from the conventional clustering coefficients, the present clustering coefficients do not confound the strength of local clustering at an ROI and the magnitude of the ROI's connectivity.

\begin{figure}[t]
\begin{center}
\includegraphics[scale=0.3]{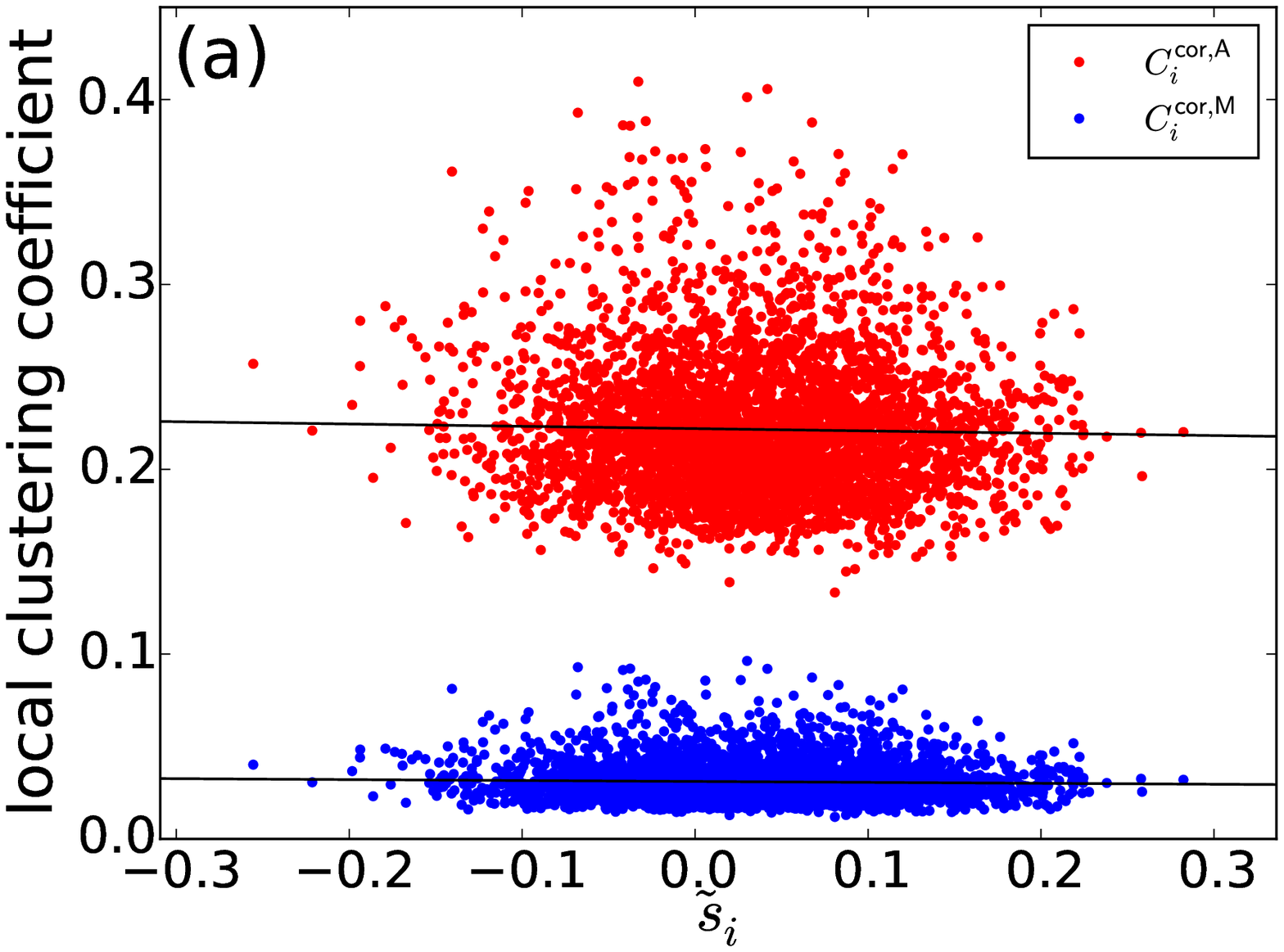}
\includegraphics[scale=0.3]{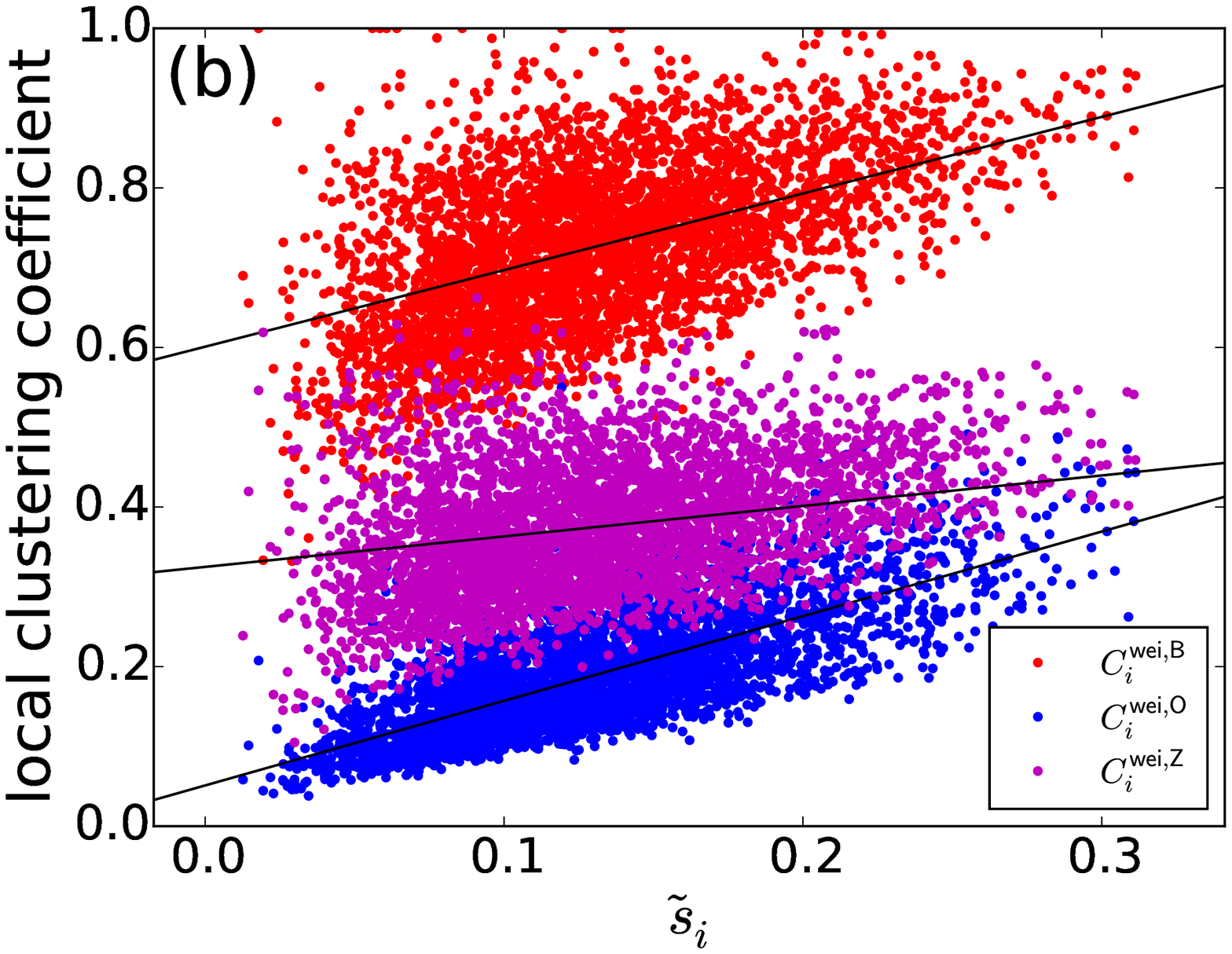}
\caption{(a) Relationship between $\tilde{s}_i$ and the local clustering coefficients for correlation matrices. (b) Relationship between $\tilde{s}_i$ and the local clustering coefficients for weighted networks. The solid lines represent the fixed effect estimated by the linear mixed model.}
\label{fig:s_i vs C_i}
\end{center}
\end{figure}

The clustering coefficients $C_i^{\rm cor,A}$ and $C_i^{\rm cor,M}$ are plotted against $\tilde{s}_i \equiv s_i/(N_{\rm ROI}-1)$, i.e., the node strength normalised between $-1$ and $1$, in 
Fig.~\ref{fig:s_i vs C_i}(a), where a symbol represents a combination of an ROI and an individual.
Figure~\ref{fig:s_i vs C_i}(a) suggests that $s_i$ and the local clustering coefficient are uncorrelated.
To statistically prove this casual observation, we fitted a linear mixed-effects model for each type of local clustering coefficient.  In the linear mixed-effects model, the local clustering coefficient value for the combination of a participant and an ROI was the dependent variable ($n=138$ participants and $N_{\rm ROI}=30$ ROIs). The independent variable was the equivalent of the node strength, i.e., $\sum_{j=1; j\neq i}^{N_{\rm ROI}} \rho(i,j)$. We assumed random effects over participants influencing the slope and intercept.
We found that $C_i^{\rm cor,A}$ and $C_i^{\rm cor,M}$ did not show strong positive correlation with $\sum_{j=1; j\neq i}^{N_{\rm ROI}} \rho(i,j)$ ($C_i^{\rm cor,A}$: $t_{4139}=-2.33$, $P=0.023$, Pearson correlation coefficient between $C_i^{\rm cor,A}$ and $\sum_{j=1; j\neq i}^{N_{\rm ROI}} \rho(i,j)$, $i=1, \ldots, N_{\rm ROI}$ for each participant, which is then averaged over all the participants, as a measure of effect size $r_{28} = -0.023$, $C_i^{\rm cor,A} = -0.013 \tilde{s}_i + 0.222$; $C_i^{\rm cor,M}$: $t_{4139}=-3.20$, $P=0.0019$, $r_{28} = -0.047$, $C_i^{\rm cor,M} = -0.0050 \tilde{s}_i + 0.031$). Note that the effect size as measured by $r_{28}$ was small, although the effects were significant owing to a large sample size.

We investigated the same linear relationship for the correlation matrices generated by the randomization of the original correlation matrices using the H-Q-S algorithm. We generated one null model network per participant. For four participants, the algorithm did not work because the average off-diagonal element of the covariance matrix for the empirical covariance matrix was negative, violating the condition for the algorithm to be used \cite{Hirschberger2007EurJOperRes}. For the remaining $n-4=134$ participants,
the dependence of the local clustering coefficient of ROI $i$ on $\sum_{j=1; j\neq i}^{N_{\rm ROI}} \rho(i,j)$ remained small ($C_i^{\rm cor,A}$: $t_{4019}= -1.93$, $P = 0.059$, $r_{28} = -0.021$, $C_i^{\rm cor,A} = -0.0051 \tilde{s}_i + 0.28$; $C_i^{\rm cor,M}$: $t_{4019}=-1.21$, $P=0.23$, $r_{28} = -0.019$, $C_i^{\rm cor,M} = -0.0016\tilde{s}_i + 0.055$). Therefore, we conclude that $C_i^{\rm cor,A}$ and $C_i^{\rm cor,M}$ (and hence $C^{\rm cor, A}$ and $C^{\rm cor,M}$) are not affected by pseudo-correlation and provide measurements orthogonal to the node strength.

In contrast, the previously provided local clustering coefficients for unweighted or weighted networks (i.e., $C_i^{\rm unw}$, $C_i^{\rm wei,B}$, $C_i^{\rm wei,O}$ and $C_i^{\rm wei,Z}$ given by Eqs.~\eqref{eq:C_i}, \eqref{eq:C_i^Barrat}, \eqref{eq:C_i^Onnela} and \eqref{eq:C_i^Zhang}, respectively) should be correlated with the degree (i.e., the number of edges connected to a node), $k_i$ (in the case of unweighted networks) or node strength, i.e., weighted degree $s_i$ (in the case of weighted networks) when applied to correlation matrices. Let us explain this point for weighted networks for the sake of clarity. Because of indirect paths, if $w_{ij}$ and $w_{i\ell}$ are large, $w_{j\ell}$ tends to be large, which increases the value of the local clustering coefficient of ROI $i$. At the same time, $s_i$ is large if $w_{ij}$ and $w_{i\ell}$ are. Therefore, we expect systematic positive correlation between $s_i$ and any of $C_i^{\rm unw}$, $C_i^{\rm wei,B}$, $C_i^{\rm wei,O}$ and $C_i^{\rm wei,Z}$.

The three types of clustering coefficient for weighted networks ($C_i^{\rm wei,B}$, $C_i^{\rm wei,O}$ and $C_i^{\rm wei,Z}$) are plotted against $\tilde{s}_i$ in Fig.~\ref{fig:s_i vs C_i}(b). We did not examine the local clustering coefficient for unweighted networks (i.e., $C^{\rm unw}_i$) because it was undefined for many ROIs, whose nodal degree $k_i$ was either 0 or 1; our network is relatively small (i.e., $N_{\rm ROI} = 30$) and the edge density is not assumed to be too large. The three local weighted clustering coefficients and $\tilde{s}_i$ were strongly correlated ($C_i^{\rm wei,B}$: $t_{4139} = 23.7$ for the fixed effects of $\tilde{s}_i$,
%
% using the \textit{lmer} function in lme4 package in R (v. 2.15.2?) [ref],
%
 $P<10^{-15}$, $r_{28} = 0.43$, the estimated fixed effects: $C_i^{\rm wei,B} = 0.960 \tilde{s}_i + 0.601$;  $C_i^{\rm wei,O}$: $t_{4139}=43.4$, $P<10^{-15}$, $r_{28} = 0.70$, $C_i^{\rm wei,O} = 0.950 \tilde{s}_i + 0.064$; $C_i^{\rm wei,Z}$: $t_{4139}=10.8$, $P<10^{-15}$, $r_{28} = 0.27$, $C_i^{\rm wei,Z} = 0.382 \tilde{s}_i + 0.325$).
 
Upon randomisation of the original correlation matrices by the H-Q-S algorithm, the positive relationship between the local clustering coefficient and $\tilde{s}_i$ persisted for each weighted clustering coefficient index ($C_i^{\rm wei,B}$: $t_{4019} = 13.1$, $P<10^{-15}$, $r_{28} = 0.27$, $C_i^{\rm wei,B} = 0.509 \tilde{s}_i + 0.595$; $C_i^{\rm wei,O}$: $t_{4019}=37.0$, $P<10^{-15}$, $r_{28} = 0.60$, $C_i^{\rm wei,O} = 0.628 \tilde{s}_i + 0.100$; $C_i^{\rm wei,Z}$: $t_{4019}=8.56$, $P=3.7\times 10^{-13}$,
$r_{28} = 0.17$, $C_i^{\rm wei,Z} = 0.217 \tilde{s}_i + 0.355$). These results suggest that these local clustering coefficients are confounded by the effect of node strength, which could arise from the pseudo-correlation due to indirect paths.

\subsection{Dependence of the local clustering coefficients on the brain system\label{sub:system dependence}}

Previous studies found systematic regional differences (e.g., across different lobes) in the local clustering coefficient in functional brain networks
\cite{Achard2006JNeurosci,Alexanderbloch2010FrontSystNeurosci,Lynall2010JNeurosci,Salallonch2014NeurobiolAging}.
However, this effect may be confounded by the effect of the node strength. As a case study, in this section we show that we do not see the association between previously defined brain systems (i.e., subsets of the ROIs constituting the entire network) and age-related changes in conventional local clustering coefficients if the effect of the node strength is controlled.

We first calculated the Pearson correlation coefficient ($r$) between the age and a nodal index such as the local clustering coefficient at each ROI. Then, we examined whether $r$ was different across three brain systems whose functions and structures have been examined \cite{Fair2009PlosComputBiol,Power2011Neuron}: the default mode network (DMN), cingulo-opercular network (CON) and fronto-parietal network (FPN).

\begin{figure}[t]
\begin{center}
\includegraphics[scale=0.4]{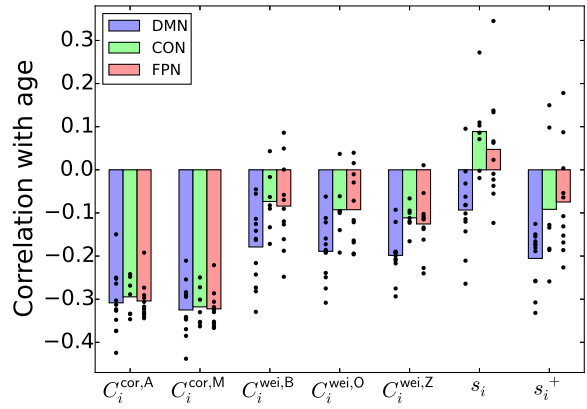}
\caption{Pearson correlation coefficient between a nodal index and the age, averaged over the ROIs in the DMN, CON or FPN. The circle represents the correlation coefficient value for a single node.}
\label{fig:local vs system}
\end{center}
\end{figure}

The $r$ values between various nodal indices and the age, averaged over the ROIs in each of the DMN, CON and FPN, are shown in Fig.~\ref{fig:local vs system}. For the clustering coefficients for weighted networks (i.e., $C^{\rm wei,B}$, $C^{\rm wei,O}$ and $C^{\rm wei,Z}$), $r$ was negative for most ROIs, confirming the results reported in section~\ref{sub:corr with age and exec} that the (global) clustering coefficient was negatively correlated with the age of the participant. The $r$ value was different between the three brain systems for each type of weighted clustering coefficient
($C^{\rm wei,B}_i$: $F_{2,27} = 4.32$, $P=0.023$, $\eta^2=0.24$; $C^{\rm wei,O}_i$: $F_{2,27}=5.69$, $P=0.0087$, $\eta^2=0.30$; $C^{\rm wei,Z}$: $F_{2,27}=6.87$, $P=0.0039$, $\eta^2=0.34$; a one-way factorial analysis of variance (ANOVA) [System: DMN/CON/FPN]). Post-hoc two-sample $t$-tests revealed that the effect of the age was larger in the DMN than in the CON and FPN ($C^{\rm wei,B}_i$, DMN $-$ CON: $t_{17} = -2.64$, $P=0.017$, $d=-1.28$; 
$C^{\rm wei,B}_i$, DMN $-$ FPN: $t_{21} = -2.38$, $P=0.017$, $d=-1.04$; 
$C^{\rm wei,O}_i$, DMN $-$ CON: $t_{17} = -2.86$, $P=0.011$, $d=-1.39$; 
$C^{\rm wei,O}_i$, DMN $-$ FPN: $t_{21} = -2.95$, $P=0.00077$, $d=-1.29$; 
$C^{\rm wei,Z}_i$, DMN $-$ CON: $t_{17} = -3.84$, $P=0.0013$, $d=-1.86$; 
$C^{\rm wei,Z}_i$, DMN $-$ FPN: $t_{21} = -2.78$, $P=0.011$, $d=-1.21$).

However, qualitatively the same association between the age and the brain system was also found when $r$ was defined as the correlation between the node strength (i.e., $s_i$) and the age
($F_{2,27} = 8.01$, $P=0.0019$, $\eta^2=0.37$) and when $r$ was defined as the correlation between $s_i^+$, which was defined as $\sum_{j=1; \rho(i,j)> 0}^{N_{\rm ROI}} \rho(i,j)$, and the age ($F_{2,27} = 4.43$, $P=0.022$, $\eta^2=0.25$). Because the local clustering coefficients for weighted networks (i.e., $C^{\rm wei,B}_i$, $C^{\rm wei,O}_i$ and $C^{\rm wei,Z}_i$) were positively correlated with the node strength and $s_i^+$, we take $s_i$ or $s_i^+$ as a simpler signature of the system dependence of the age effect than the local clustering coefficient. 

In contrast, the proposed local clustering coefficients, which were not correlated with $s_i$ or $s_i^+$ (Fig.~\ref{fig:s_i vs C_i}(a)), were not different across the brain systems ($C^{\rm cor,A}_i$: $F_{2,27} = 0.13$, $P=0.88$, $\eta^2=0.01$; $C^{\rm cor,M}_i$: $F_{2,27} = 0.04$, $P=0.96$, $\eta^2=0.003$; also see Fig.~\ref{fig:local vs system}). These observations suggest that the apparent dependence of the clustering coefficient on the brain system when a conventional clustering coefficient is used is explained by the nodal measure, $s_i$ or $s_i^+$.
 
We found similar results in sensory-motor regions in the brain (Appendix~C). In other words, the association between the clustering coefficient and the age is more positive for the ROIs in a somatosensory-motor system than for the ROIs in an auditory system and a visual system when we used the clustering coefficients for weighted networks. Qualitatively the same dependence on the brain system was also found when we looked at the association between the node strength and the age. In contrast, with the proposed local clustering coefficients, the auditory system showed the strongest association between the clustering coefficient and the age. These results bear robustness to our suggestion that the proposed local clustering coefficients are not confounded by the node's strength, whereas the conventional clustering coefficients are.

\section{Discussion}

We proposed two clustering coefficients tailored to correlation matrices. They do not suffer from pseudo-correlation induced by indirect paths between two ROIs through a third ROI, do not require thresholding, do not discard negative pairwise correlation, and do not suffer from the difficulty in estimating partial correlation matrices. The proposed clustering coefficients were more strongly correlated with the participants' age than the conventional clustering coefficients, including those calculated for partial correlation matrices, were. In addition, our clustering coefficients can be used as a local measure to characterise nodes, whereas the counterparts for the conventional clustering coefficients were confounded with the (weighted) degree of the node. These results hold true for two alternative definitions of the clustering coefficient for correlation matrices that we additionally propose (Appendix~A).

Previous research has produced incongruent results regarding the changes in the clustering coefficient along ageing. In an fMRI study, both at rest and during tasks, the clustering coefficient in functional networks decreased with ageing \cite{Grady2016NeurobiolAging}, which is consistent with the present results. This observation is also consistent with results of an EEG study at rest \cite{Knyazev2015NeurobiolAging}. In different studies, however, no difference was found in the clustering coefficient between younger and older individuals \cite{WangLi2010Neuroimage,Brier2014NeurobiolAging}, or the clustering coefficient increased with ageing \cite{Matthaus2012BrainConn,ZhuWen2012NeurobiolAging,LiuKe2014PlosOne,Salallonch2014NeurobiolAging}. The diversity in these results may owe to participant's heterogeneity, inefficiency of the conventional clustering coefficients or other reasons. It should be noted that the decrease in the clustering coefficient found in a recent study \cite{Grady2016NeurobiolAging} and the present study is consistent with the decline in small-worldness of brain networks, which have been documented by using different indices \cite{Achard2007PlosComputBiol,GongRosaneto2009JNeurosci}.
However, we do not claim that the decline in the clustering coefficient along ageing is a ground truth. In fact, the coordinates of the ROIs in the current data set were determined from participants aged 7--31 \cite{Fair2009PlosComputBiol} so that they may not reflect functional ROIs in older adults \cite{ChanPark2014PNAS,Geerligs2017HumanBrainMapping}. This issue warrants further study.

We demonstrated the utility of the proposed correlation coefficients with fMRI data collected from individuals of different ages. They may also be useful in deciphering functional brain networks collected from different types of individuals such as those with psychiatric or other disorders, those under different task conditions and children under developments. Furthermore, the present method can be used to any correlation or covariance matrix, thus promising their applicability to other functional data of the brain, such as EEG, MEG, correlation in the cortical thickness between ROIs, where correlation is calculated across individuals (see Introduction for references), and even correlation data outside neuroscience. 

The proposed clustering coefficients are expected to find immediate applications in the assessment of small-worldness. In the small-world analysis, a major method is to combine the clustering coefficient and the average path length between a pair of nodes, denoted by $L$. When $L$ is small and the clustering coefficient is large, one says that the network is small-world \cite{Watts1998Nature,Bullmore2009NatRevNeurosci} (but see \cite{Achard2007PlosComputBiol,GongRosaneto2009JNeurosci} for different definitions based on the so-called network efficiency indices). In neuroscience, it is often the case to combine these two indices to examine a single small-worldness index \cite{Humphries2006ProcRSocB} (also see \cite{Muldoon2016SciRep} for a recent development). The motivation behind the present study is that the definition or measurement of clustering is nontrivial for correlation matrices, i.e., functional data.

The same caution applies to the path length. A common way to calculate the path length in correlation data is to threshold on the correlation matrix to generate an unweighted network and then measure the path length. However, this method suffers from arbitrariness of thresholding, as discussed in Introduction. Another common way is to define a relationship between the edge weight, i.e., correlation coefficient value, and the cost of passing through the edge. Popular choices of the cost function are the reciprocal of the edge weight \cite{Rubinov2010Neuroimage} and a constant subtracted by the edge weight \cite{Achard2007PlosComputBiol,GongRosaneto2009JNeurosci}. However, the theoretical basis of these decisions seems unclear. A more sensible definition of the distance between ROIs $i$ and $j$ may be $\sqrt{2(1-\rho(i,j))}$, which qualifies as a Euclidean distance \cite{Mantegna2000book}.

We used the three-way partial correlation coefficient controlling for a single ROI to define the clustering coefficients. In contrast, some previous studies derived functional networks from partial correlation matrices \cite{Salvador2005CerebCortex,Marrelec2006Neuroimage,SmithMiller2011Neuroimage}. Both types of methods intend to remove the spurious correlation induced by indirect paths between ROIs. While getting common, the methods based on partial correlation matrices face a technical challenge that the partial correlation matrix cannot be determined uniquely from data in general \cite{Schafer2005StatApplGenetMolBiol,Ryali2012Neuroimage,Brier2015Neuroimage}. In addition, its calculation for a single pair of nodes involves all the other $N_{\rm ROI}-2$ nodes, contradicting the original premise of the clustering coefficient that it is a local quantity \cite{Watts1998Nature}. Our clustering coefficients, which use the three-way partial correlation coefficient, do not suffer from the non-uniqueness problem and is a local quantity. Furthermore, we showed that the present clustering coefficients were associated with the age, whereas those calculated for the partial correlation matrices were not.
Generalisation of this finding to different ROIs, data sets and types of participants, such as those with a particular brain-related disorder, warrants future work.

\section*{Appendix~A: Two alternative clustering coefficients for correlation matrices}

We assessed two alternative clustering coefficients for correlation matrices on the empirical and randomised data.

In the first variant, we restricted ourselves to the cases in which $\rho(i,j)$ and $\rho(i,\ell)$ were positive when calculating the local clustering coefficient at ROI $i$, denoted by $C_i^{\rm cor,P}$ (superscript P standing for positive). We set
\begin{equation}
C_i^{\rm cor,P} \equiv \frac{\sum_{\substack{1\le j< \ell\le N_{\rm ROI}\\ j,\ell\neq i; \rho(i,j),\rho(i,\ell)>0}} \rho(i,j) \rho(i,\ell) \rho^{\rm partial}(j,\ell \mid i)}
{\sum_{\substack{1\le j< \ell\le N_{\rm ROI}\\ j,\ell\neq i; \rho(i,j),\rho(i,\ell)>0}} \rho(i,j) \rho(i,\ell)}.
\label{eq:C_i^P}
\end{equation}
In other words, $C_i^{\rm cor,P}$ is a weighted average of the partial correlation over pairs of $j$ and $\ell$ ($j, \ell \neq i, j\neq \ell$) for which $\rho(i, j), \rho(i, \ell) >0$.  The corresponding global clustering coefficient, denoted by $C^{\rm cor,P}$, is given by the average of $C^{\rm cor,P}_i$ over all nodes.
Note that $-1\le C_i^{\rm cor,P}\le 1$ $(1\le i\le N_{\rm ROI})$ and that $-1\le C^{\rm cor,P}\le 1$.

The second variant of the clustering coefficient uses contributions of all ROIs regardless of the sign of $\rho(i,j)$ and $\rho(i,\ell)$, but in a different manner from $C_i^{\rm cor,A}$ (and hence $C^{\rm cor,A}$). If $\rho(i,j),\rho(i,\ell) < 0$, Eq.~\eqref{eq:def partial cor} implies that $\rho(j,\ell)$ would be positive if there is no partial correlation between $j$ and $\ell$. Therefore, we regard that $\rho^{\rm partial}(j, \ell \mid i)$ measures the excess correlation between $j$ and $\ell$ as usual. In contrast, if $\rho(i,j)\rho(i,\ell) < 0$, Eq.~\eqref{eq:def partial cor} implies that $\rho(j,\ell)$ would be negative if there is no partial correlation between $j$ and $\ell$. This observation is consistent with Heider's balance theory, originating from social psychology and respected in various signed network data, which dictates that in signed unweighted networks (edge weight is either $+1$ or $-1$), triangles with one or three $+1$'s are stable, whereas those with zero or two $+1$'s are unstable  \cite{Heider1946JPsych,Wasserman1994book,Szell2010PNAS,Facchetti2011PNAS}. A related remark was previously made for correlation matrices \cite{Costantini2014PlosOne}. When $\rho(i,j)\rho(i,\ell) < 0$, we regard that $\rho(j,\ell)$ being more negative (towards $-1$) than $\rho(i,j)\rho(i,\ell)$ is a signature of strong association between ROIs $j$ and $\ell$ with the influence of ROI $i$ controlled. In other words, we take a negative large value of $\rho^{\rm partial}(j, \ell \mid i)$ as an indication of  clustering composed of the three ROIs, $i$, $j$ and $\ell$, from the viewpoint of $i$. On the basis of this reasoning, we define the local clustering coefficient denoted by $C_i^{\rm cor,H}$ (superscript H standing for Heider) as follows:
\begin{equation}
C_i^{\rm cor,H} = \frac{\sum_{\substack{1\le j< \ell\le N_{\rm ROI}\\ j,\ell\neq i}} \rho(i,j) \rho(i,\ell) \rho^{\rm partial}(j,\ell \mid i)}
{\sum_{\substack{1\le j< \ell\le N_{\rm ROI}\\ j,\ell\neq i}} \left|\rho(i,j) \rho(i,\ell)\right|}.
\label{eq:C_i^H}
\end{equation}
The denominator on the right-hand side of Eq.~\eqref{eq:C_i^H} is for normalisation to guarantee $-1\le C_i^{\rm cor,H} \le 1$ $(1\le i\le N)$. The corresponding global clustering coefficient, denoted by $C^{\rm cor,H}$, is given by the average of $C^{\rm cor,H}_i$ over all nodes. Note that $-1\le C^{\rm cor,H}\le 1$. We also note that $C_i^{\rm cor,P}$, $C^{\rm cor,P}$, $C_i^{\rm cor,H}$ and $C^{\rm cor,H}$ can be negative, which is different from the clustering coefficients for unweighted and weighted networks and also from $C_i^{\rm cor,A}$, $C^{\rm cor,A}$, $C_i^{\rm cor,M}$ and $C^{\rm cor,M}$.

Both $C^{\rm cor,P}$ and $C^{\rm cor,H}$ were larger for the empirical data than for white-noise signals (for $C^{\rm cor,P}$, empirical: mean $\pm$ sd $= 0.109 \pm 0.038$, white noise: $0.0003 \pm 0.0044$, $t_{137}=33.4$, $P<10^{-6}$, $d=5.71$; for $C^{\rm cor,H}$, empirical: $0.090 \pm 0.040$, white noise: $-0.0001 \pm 0.0017$, $t_{137}=26.7$, $P<10^{-6}$, $d=4.57$). This result is consistent with that for $C^{\rm cor,A}$ and $C^{\rm cor,M}$. In addition, the empirical $C^{\rm cor,P}$ and $C^{\rm cor,H}$ values were larger than those for randomised signals generated by the H-Q-S algorithm (for $C^{\rm cor,P}$, H-Q-S: mean $\pm$ sd $= 0.035 \pm 0.032$, $t_{133}=22.0$, $P<10^{-6}$, $d=3.81$; for $C^{\rm cor,H}$, H-Q-S: $0.003 \pm 0.017$, $t_{133}=25.8$, $P<10^{-6}$, $d=4.47$). This result is opposite to  that for $C^{\rm cor,A}$ and $C^{\rm cor,M}$. In sum, $C^{\rm cor,P}$ and $C^{\rm cor,H}$ were larger for the empirical data than for both types of randomised data.

Similar to $C^{\rm cor, A}$ and $C^{\rm cor, M}$, we found that $C^{\rm cor,P}$ and $C^{\rm cor,H}$ were strongly correlated with $s^+$ ($C^{\rm cor,P}$: $r_{136} = 0.874$, $P<10^{-6}$; $C^{\rm cor,H}$: $r_{136}=0.773$, $P<10^{-6}$) but not with $s$ ($C^{\rm cor,P}$: $r_{136} = 0.197$, $P=0.020$; $C^{\rm cor,H}$: $r_{136} = -0.104$, $P=0.23$). The correlation between these clustering coefficients and the age was moderate ($C^{\rm cor,P}$: $r_{136} = -0.310$, $P=2.1\times 10^{-4}$; $C^{\rm cor,H}$: $r_{136} = -0.329$, $P<8.3\times 10^{-5}$) but was not significant when we control for the effect of $s^+$
($C^{\rm cor,P}$: $r_{136} = -0.083$, $P=0.33$; $C^{\rm cor,H}$: $r_{136} = -0.146$, $P=0.087$).

To conclude, the results regarding the association of $C^{\rm cor,P}$ and $C^{\rm cor,H}$ with the age are consistent with but weaker than those for $C^{\rm cor,A}$ and $C^{\rm cor,M}$.

Local clustering coefficients $C_i^{\rm cor,P}$ and $C_i^{\rm cor,H}$ did not show strong positive correlation with the equivalent of the node strength, i.e., $\sum_{j=1; j\neq i}^{N_{\rm ROI}} \rho(i,j)$ ($C_i^{\rm cor,P}$: $t_{4139}=-1.58$, $P=0.12$, $r_{28} = -0.057$, $C_i^{\rm cor,P} = -0.043 \tilde{s}_i + 0.110$; $C_i^{\rm cor,H}$: $t_{4139}=-2.86$, $P=0.0053$, $r_{28} = -0.075$, $C_i^{\rm cor,H} = -0.033 \tilde{s}_i + 0.091$). Note that the Pearson correlation coefficient values were small, whereas the $t$ value was significant due to a large sample size. Upon randomisation, the dependence of the local clustering coefficient of ROI $i$ on $\sum_{j=1; j\neq i}^{N_{\rm ROI}} \rho(i,j)$ remained small in terms of the $r$ value ($C_i^{\rm cor,P}$: $t_{4019}= -5.14$, $P=1.0\times 10^{-6}$,
$r_{28} = -0.19$, $C_i^{\rm cor,P} = -0.110 \tilde{s}_i + 0.038$; $C_i^{\rm cor,H}$: $t_{4019}=0.746$, $P=0.46$, $r_{28} = 0.011$, $C_i^{\rm cor,H} = 0.005\tilde{s}_i + 0.000$). Therefore, we conclude that $C_i^{\rm cor,P}$ and $C_i^{\rm cor,H}$, and hence $C^{\rm cor, P}$ and $C^{\rm cor,H}$ also, are not affected by indirect correlation and provide measurements orthogonal to the node strength.

\section*{Appendix~B: Difference between the proposed and conventional clustering coefficients in terms of their association with the age}

Table~\ref{tab:corr with age} suggests that the correlation between the proposed clustering coefficients and the age is stronger than that between the conventional clustering coefficients and the age. To examine this point statistically, we ran the Williams' $t$-test for two non-independent correlation coefficients with a variable in common \cite{Weaver2013BehavResMeth}. The common variable was the age. We compared each of the two proposed clustering coefficients, $C^{\rm cor,A}$ and $C^{\rm cor,M}$, with each of the five conventional clustering coefficients, $C^{\rm unw}$ with edge density $0.1$,
$C^{\rm unw}$ with edge density $0.2$,
$C^{\rm wei, B}$, $C^{\rm wei,O}$ and $C^{\rm wei,Z}$.
The results ($C^{\rm cor,A}$ vs $C^{\rm unw}$ with edge density $0.1$: $t_{135} = -1.93$, $p = 0.028$;
$C^{\rm cor,A}$ vs $C^{\rm unw}$ with edge density $0.2$: $t_{135} = -2.58$, $p = 0.0055$;
$C^{\rm cor,A}$ vs $C^{\rm wei, B}$: $t_{135} = -1.86$, $p = 0.033$;
$C^{\rm cor,A}$ vs $C^{\rm wei,O}$: $t_{135} = -3.09$, $p = 0.0012$;
$C^{\rm cor,A}$ vs $C^{\rm wei,Z}$: $t_{135} = -3.03$, $p = 0.0015$;
$C^{\rm cor,M}$ vs $C^{\rm unw}$ with edge density $0.1$: $t_{135} = -2.26$, $p = 0.013$;
$C^{\rm cor,M}$ vs $C^{\rm unw}$ with edge density $0.2$: $t_{135} = -2.89$, $p = 0.0022$;
$C^{\rm cor,M}$ vs $C^{\rm wei, B}$: $t_{135} = -2.14$, $p = 0.017$;
$C^{\rm cor,M}$ vs $C^{\rm wei,O}$: $t_{135} = -3.10$, $p = 0.0012$;
$C^{\rm cor,M}$ vs $C^{\rm wei,Z}$: $t_{135} = -3.08$, $p = 0.0013$;
all $p$ values were not corrected for multiple comparison) indicate that seven out of the ten cases survived Bonferroni correction ($ \alpha = 5\%$). Therefore, we conclude that the two proposed clustering coefficients are more strongly associated with the age than the conventional clustering coefficients are.

\section*{Appendix~C: Dependence of the local clustering coefficients on sensory-motor brain systems}

We repeated the same analysis as that in section~\ref{sub:system dependence} for sensory-motor brain systems. Because the brain atlas used in the main text only has the DMN, CON, FPN and cerebellum \cite{Fair2009PlosComputBiol}, we used the somatosensory-motor network (SMN; 34 ROIs), auditory network (13 ROIs) and visual network (31 ROIs), which are among several brain systems identified in a different study \cite{Power2011Neuron}.

\begin{figure}[t]
\begin{center}
\includegraphics[scale=0.4]{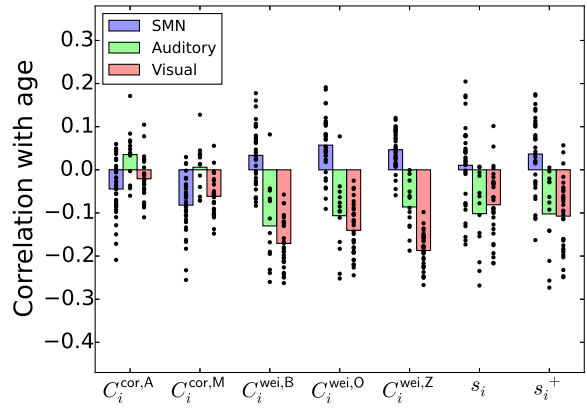}
\caption{Pearson correlation coefficient between a nodal index and the age, averaged over the ROIs in the SMN, auditory network or visual network. The circle represents the correlation coefficient value for a single node.}
\label{fig:local vs system SMN AUD VIS}
\end{center}
\end{figure}

The Pearson correlation coefficient, $r$, between the different clustering coefficients and the age and that between the node strength and the age, averaged over the ROIs in each of the SMN, auditory network and visual network, is shown in Fig.~\ref{fig:local vs system SMN AUD VIS}. For the clustering coefficients for weighted networks (i.e., $C^{\rm wei,B}$, $C^{\rm wei,O}$ and $C^{\rm wei,Z}$), $r$ was slightly positive on average in the SMN and moderately or considerably negative in the auditory and visual networks. The $r$ value was different between the three brain systems for each type of weighted clustering coefficient
($C^{\rm wei,B}_i$: $F_{2,75} = 72.6$, $P<10^{-15}$, $\eta^2=0.66$; $C^{\rm wei,O}_i$: $F_{2,75}=72.0$, $P<10^{-15}$, $\eta^2=0.66$; $C^{\rm wei,Z}$: $F_{2,75}=211$, $P<10^{-15}$, $\eta^2=0.85$; a one-way ANOVA [System: SMN/Auditory/Visual]). Post-hoc two-sample $t$-tests revealed that the effect of the age was more positive in the SMN than in the auditory and visual networks ($C^{\rm wei,B}_i$, SMN $-$ Auditory: $t_{45} = 6.37$, $P=8.9\times 10^{-9}$, $d=1.90$; 
$C^{\rm wei,B}_i$, SMN $-$ Visual: $t_{63} = 12.8$, $P<10^{-15}$, $d=3.23$; 
$C^{\rm wei,O}_i$, SMN $-$ Auditory: $t_{45} = 6.79$, $P=2.1\times 10^{-9}$, $d=2.03$; 
$C^{\rm wei,O}_i$, SMN $-$ Visual: $t_{63} = 12.3$, $P<10^{-15}$, $d=3.10$; 
$C^{\rm wei,Z}_i$, SMN $-$ Auditory: $t_{45} = 8.30$, $P=1.3\times 10^{-11}$, $d=2.47$; 
$C^{\rm wei,Z}_i$, SMN $-$ Visual: $t_{63} = 21.8$, $P<10^{-15}$, $d=5.49$).
However, as is observed in Fig.~\ref{fig:local vs system SMN AUD VIS}, qualitatively the same association between the age and the brain system was also found when $r$ was defined as the correlation between $s_i$ and the age ($F_{2,75} = 11.6$, $P=4.2\times 10^{-5}$, $\eta^2=0.92$) and when $r$ was defined as the correlation between $s_i^+$ and the age ($F_{2,75} = 25.5$, $P=3.6\times 10^{-9}$, $\eta^2=0.96$).

The proposed local clustering coefficients were also different across the brain systems ($C^{\rm cor,A}_i$: $F_{2,75} = 9.06$, $P=0.00030$, $\eta^2=0.90$; $C^{\rm cor,M}_i$: $F_{2,75} = 11.1$, $P=6.2\times 10^{-5}$, $\eta^2=0.92$). However, as suggested in Fig.~\ref{fig:local vs system}, the brain system that showed the most positive correlation with the age was the auditory network
($C^{\rm wei,A}_i$, Auditory $-$ SMN: $t_{45} = 3.82$, $P=0.00041$, $d=1.14$; 
$C^{\rm wei,A}_i$, Auditory $-$ Visual: $t_{42} = 3.34$, $P=0.0018$, $d=1.03$; 
$C^{\rm wei,M}_i$, Auditory $-$ SMN: $t_{45} = 4.20$, $P=0.00013$, $d=1.25$; 
$C^{\rm wei,M}_i$, Auditory $-$ Visual: $t_{42} = 4.17$, $P=0.00015$, $d=1.29$). 
These results are consistent with those shown in section~\ref{sub:system dependence}.
In other words, the assoiation between the clustering coefficients for weighted networks and the age is confounded by that between the node strength and the age. In contrast, the proposed clustering coefficients measure the effect of local clustering on the age without being confounded by the node strength.

\section*{Acknowledgments}

We thank Koji Oishi for valuable feedback on the manuscript. N.M. acknowledges the support provided through JST CREST Grant Number JPMJCR1304 and the JST ERATO Grant Number JPMJER1201, Japan. M.S. acknowledges the support provided through European Commission (CIG618600) and Japan Society for the Promotion of Science (16H05959, 16KT0002 and 16Ｈ02053).
T.E. acknowledges the support provided through PRESTO, JST (No. JPMJPR16D2).

\section*{References}

%\bibliography{../../social/citations}

\newpage
\clearpage

\newpage
\clearpage

\newpage
\clearpage

\clearpage

\clearpage

\newpage
\clearpage

\begin{table}[htbp]
  \centering
\caption{Correlation between the clustering coefficient and age. The correlation coefficient is denoted by $r$. The degree of freedom is equal to $n-2=136$.}
\label{tab:corr with age} 
\begin{tabular}{lYYYY}
    \hline
    \hline
    \multicolumn{1}{c}{\multirow{2}[4]{*}{Index}} & \multicolumn{2}{c}{Unconditional} & \multicolumn{2}{c}{Effect of $s^+$ controlled} \bigstrut\\
\cline{2-5}          & $r$   & $P$   & $r$   & $P$ \bigstrut\\
    \hline
    Pearson correlation matrix &       &       &       &  \\
    \quad$C^{\rm cor,A}$ & $-0.377$ & $<10^{-5}$ & $-0.224$ & $0.0076$ \\
    \quad$C^{\rm cor,M}$ & $-0.397$ & $<10^{-5}$ & $-0.259$ & $0.0019$ \\
    \quad$C^{\rm unw}$, edge density $= 0.1$ & $-0.234$ & $0.0058$ & $-0.104$ & $0.23$ \\
    \quad$C^{\rm unw}$, edge density $=0.2$ & $-0.197$ & $0.021$ & $-0.032$ & $0.71$ \\
    \quad$C^{\rm wei, B}$ & $-0.262$ & $0.0019$ & $0.018$ & $0.83$ \\
    \quad$C^{\rm wei,O}$ & $-0.240$ & $0.0045$ & $0.014$ & $0.87$ \\
    \quad$C^{\rm wei,Z}$ & $-0.229$ & $0.0068$ & $-0.032$ & $0.71$ \\
          &       &       &       &  \\
    Partial correlation matrix &       &       &       &  \\
    \quad$C^{\rm unw}$, edge density $= 0.1$ & $-0.001$ & $0.99$ & $0.037$ & $0.67$ \\
    \quad$C^{\rm unw}$, edge density $=0.2$ & $0.048$ & $0.58$ & $0.028$ & $0.75$ \\
    \quad$C^{\rm wei, B}$ & $-0.056$ & $0.51$ & $-0.022$ & $0.80$ \\
    \quad$C^{\rm wei,O}$ & $0.057$ & $0.50$ & $0.094$ & $0.27$ \\
    \quad$C^{\rm wei,Z}$ & $0.057$ & $0.51$ & $0.076$ & $0.37$ \\
          &       &       &       &  \\
    Average connectivity &       &       &       &  \\
    \quad$s$   & $0.020$ & $0.82$ &  ---  &  --- \\
    \quad$s^+$ & $-0.311$ & $0.0002$ &  ---  &  --- \bigstrut[b]\\
    \hline
    \hline
    \end{tabular}
\end{table}

\newpage
\clearpage

\begin{table}[htbp]
  \centering
\caption{Correlation between the clustering coefficient and the node strength. The degree of freedom is equal to $n-2=136$.}
\label{tab:corr with node strength}
    \begin{tabular}{lYYYY}
    \hline
    \hline
    \multicolumn{1}{c}{\multirow{2}[4]{*}{Index}} & \multicolumn{2}{c}{$s$} & \multicolumn{2}{c}{$s^+$} \bigstrut\\
\cline{2-5}          & $r$   & $P$   & $r$   & $P$ \bigstrut\\
    \hline
     Pearson correlation matrix &       &       &       &  \\
    \quad$C^{\rm cor,A}$ & $-0.096$ & $0.26$ & $0.812$ & $<10^{-15}$ \\
    \quad$C^{\rm cor,M}$ & $-0.084$ & $0.33$ & $0.798$ & $<10^{-15}$ \\
    \quad$C^{\rm unw}$, edge density $= 0.1$ & $0.001$ & $0.99$ & $0.471$ & $<10^{-8}$ \\
    \quad$C^{\rm unw}$, edge density $=0.2$ & $0.050$ & $0.56$ & $0.550$ & $<10^{-11}$ \\
    \quad$C^{\rm wei, B}$ & $0.359$ & $<10^{-4}$ & $0.869$ & $<10^{-15}$ \\
    \quad$C^{\rm wei,O}$ & $0.022$ & $0.80$ & $0.798$ & $<10^{-15}$ \\
    \quad$C^{\rm wei,Z}$ & $-0.080$ & $0.35$ & $0.664$ & $<10^{-15}$ \\
          &       &       &       &  \\
    Partial correlation matrix &       &       &       &  \\
    \quad$C^{\rm unw}$, edge density $= 0.1$ & $0.021$ & $0.81$ & $0.115$ & $0.18$ \\
    \quad$C^{\rm unw}$, edge density $=0.2$ & $-0.097$ & $0.26$ & $-0.070$ & $0.42$ \\
    \quad$C^{\rm wei, B}$ & $0.080$ & $0.35$ & $0.113$ & $0.19$ \\
    \quad$C^{\rm wei,O}$ & $-0.006$ & $0.94$ & $0.100$ & $0.24$ \\
    \quad$C^{\rm wei,Z}$ & $-0.041$ & $0.64$ & $0.050$ & $0.56$ \bigstrut[b]\\
    \hline
    \hline
    \end{tabular}
\end{table}

\end{document}